\documentclass[aps,pra,twocolumn,showpacs]{revtex4}
\usepackage{bm}
\usepackage{ulem}
\usepackage{epsfig}
\usepackage{graphicx}
\usepackage{amssymb,amsmath,amsbsy,amsgen,amsfonts}    
\usepackage{dcolumn}
\usepackage{amsthm}
\usepackage{mathrsfs}
\usepackage{latexsym}
\usepackage{array}
\usepackage{amstext}
\allowdisplaybreaks[1]
\usepackage{txfonts}

\newcommand{\braket}[2]{\langle#1|#2\rangle}
\newcommand{\bra}[1]{\left\langle{#1}\right\vert}
\newcommand{\ket}[1]{\left\vert{#1}\right\rangle}

\newcommand{\be}{\begin{equation}}
\newcommand{\ee}{\end{equation}}
\newcommand{\ba}{\begin{array}}
\newcommand{\ea}{\end{array}}
\newcommand{\bqa}{\begin{eqnarray}}
\newcommand{\eqa}{\end{eqnarray}}
\newcommand{\pro}[3]{\left\vert#1\rangle_{#2}\langle#3\right\vert}

\DeclareSymbolFont{symbols}{OMS}{cmsy}{m}{n}

\begin{document}

\title{Quantum plasmonics with a metal nanoparticle array} 

\author{Changhyoup Lee,$^{1,2,3}$ Mark Tame,$^4$ James Lim$^{1,2,3}$  and Jinhyoung Lee$^{1,2}$ } 
\affiliation{
$^1$Department of Physics, Hanyang University, Seoul 133-791, Korea \\
$^2$Center for Macroscopic Quantum Control, Seoul National University, Seoul 151-742, Korea \\
$^3$Research Institute for Natural Sciences, Hanyang University, Seoul 133-791, Korea \\
$^4$QOLS,\,The\,Blackett\,Laboratory,\,Imperial\,College\,London,\,Prince\,Consort\,Road,\,SW7\,2BW,\,United Kingdom
}
\date{\today}

\begin{abstract}
We investigate an array of metal nanoparticles as a channel for nanophotonic quantum communication and the generation of quantum plasmonic interference. We consider the transfer of quantum states, including single-qubits as plasmonic wavepackets, and highlight the necessity of a quantum mechanical description by comparing the predictions of quantum theory with those of classical electromagnetic theory. The effects of loss in the metal are included, thus putting our investigation into a practical setting and enabling the quantification of the performance of realistic nanoparticle arrays as plasmonic quantum channels. We explore the interference of single plasmons, finding nonlinear absorption effects associated with the quantum properties of the plasmon excitations. This work highlights the benefits and drawbacks of using nanophotonic periodic systems for quantum plasmonic applications, such as quantum communication, and the generation of quantum interference. 
\end{abstract}


\maketitle

\section{Introduction}

The field of quantum plasmonics is currently experiencing intense interest from the plasmonics and quantum optics communities~\cite{Alte,plasmonQIP,Fasel,Lukin1,Lukin2,TSPP,DSPP,DSPP2,Lukin3,Lin,Kol,Fedutik,Chang3,Berg,Anders,Cuche,Heeres,Gonz,Martin,Chenx,Huck2,Rup,Dzsot,DiMartino}. Integrated quantum systems featuring surface plasmons are showing remarkable potential for their use in quantum control applications, such as quantum information processing~\cite{Lukin2,Lin,Cuche,Heeres,DiMartino}. Here, novel capabilities in the way the electromagnetic field can be localized~\cite{Tak,Gramotnev} and manipulated~\cite{Zayats,photoncircuit,Maier2,Pendry, WasserShan} offer the prospect of miniaturization, scalability and strong coherent coupling to single emitter systems that conventional photonics cannot achieve~\cite{Lukin1,Lukin2,Kol,Fedutik,Chang3,Berg,Dzsot,Huck2,Rup}. Recent studies have focused on entanglement preservation~\cite{Alte, Fasel}, quadrature-squeezed surface plasmon propagation~\cite{Anders} and the use of surface plasmons as mediators of entanglement between two qubits~\cite{Gonz,Martin,Chenx}. With the advancement of nanofabrication techniques, ordered arrays of closely spaced noble metal nanoparticles have been proposed as a means of guiding electromagnetic energy, via localised surface plasmons (LSPs), on scales far below the diffraction limit~\cite{Quinten,Maierexp}. Here, energy transport relies on near-field coupling between surface plasmons of neighbouring particles~\cite{Brong}, with the suppression of radiative scattering into the far-field~\cite{Krenn, Krenn2, Maier}. Recently it was shown that an appropriate arrangement of nanoparticles can form passive linear nanoscale optical devices such as beam splitters, phase shifters and crossover splitters~\cite{YK,Kub,Baer}. While much progress has been made in the area of device design, so far there has been no analysis of the effects of loss in these nanoparticle systems in the quantum regime. It is vital to understand the impact of these effects on the performance of such devices so that plasmonic systems may be developed as an efficient platform for nanophotonic quantum control applications.

In this work we carry out such an analysis and investigate quantum state transfer and interference of surface plasmons on a metal nanoparticle array. The transfer of quantum states, including those encoded into single-qubit plasmon wavepackets, is studied. The effects of loss in the metal due to electronic relaxation are also included in our model, putting the investigation into a more practical setting. We find that quantum state transfer can be achieved for small length arrays even under nonideal conditions and therefore these arrays may act as channels for short distance on-chip nanophotonic quantum communication. We also study the interference of single plasmons in the nanoparticle array and find nonlinear absorption effects associated with the quantum properties of the plasmon excitations. Our study highlights the benefits and drawbacks associated with building nanophotonic systems that use surface plasmons in the quantum regime. The results of this work may help in the future study and design of more complex plasmonic structures involving emitter systems for quantum control applications, and the probing of novel nanoscale optical phenomena.

We start our investigation in the next section by introducing the nanoparticle array model and quantized mathematical description, along with some basic properties of the system dynamics. Then in section III we study the performance of quantum state transfer under ideal conditions and highlight the necessity of a quantum mechanical description by comparing the predictions of quantum theory with those of classical electromagnetic theory. In Section IV we consider the effects of damping due to losses associated with the electronic response within the metallic nanoparticles and study the interference of single plasmons in the nanoparticle array. Finally, we summarize our findings in Section V.
\begin{figure*}[t]
\centering
\includegraphics[width=17cm]{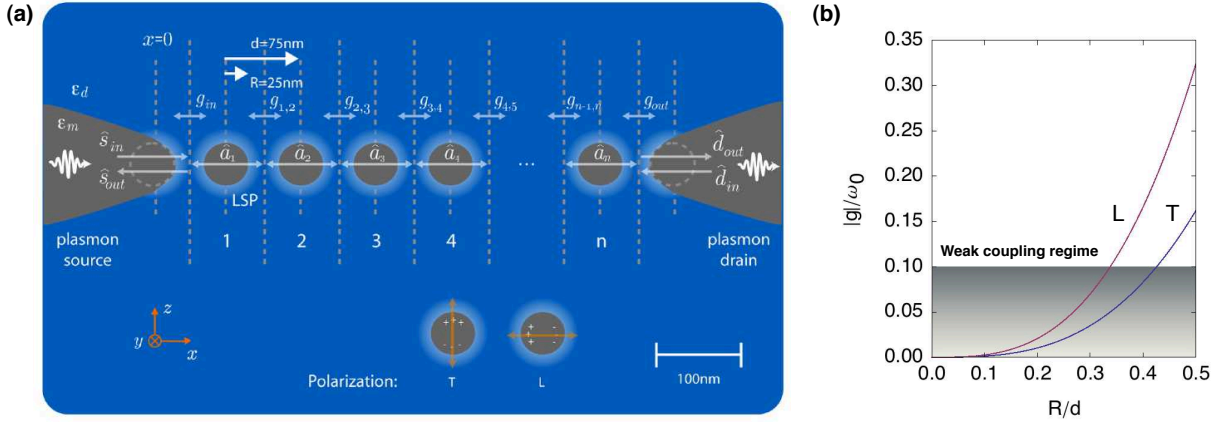}
\caption{\textbf{(a)}: A tapered metal nanowire waveguide on the left hand side focuses light to the end of its tip and excites a localised surface plasmon (LSP) on the adjacent nanoparticle. The excitation then propagates across the array of nanoparticles and exits via another tapered metal nanowire waveguide on the right hand side. All metal regions have permittivity $\epsilon_m$ and dielectric regions have permittivity $\epsilon_d$, as defined in the text. The dimensions are chosen as an example and the theory developed is more general, with a range of parameters investigated in this work. \textbf{(b)}: Weak coupling approximation for the nanoparticle array. The theoretical model developed in this paper is valid in the regime $|g| \ll \omega_{0}$ and in particular we choose ${\rm max}|g|=0.1\omega_{0}$, where $\omega_0$ is the natural frequency of the nanoparticle field oscillations and $g$ is the nearest neighbour coupling parameter. The lower blue curve is for transverse polarization (T) and upper red curve is for longitudinal (L) polarization.}
\label{figset}
\end{figure*}

\section{Physical system and Model}
\subsection{The Hamiltonian}

We consider the system depicted in Fig.~\ref{figset}~{\bf (a)}, which is presented in a top-down view. Here, a tapered metal nanowire waveguide on the left hand side focuses light at the end of its tip in the form of a confined surface plasmon field. This field then couples to the adjacent spherical metal nanoparticle and excites a localised surface plasmon (LSP). The LSP excitation propagates across the linear array of metal nanoparticles by near-field coupling and exits via another tapered metal nanowire waveguide on the right hand side. All metal regions have a frequency dependent permittivity $\epsilon_m(\omega)$ and dielectric regions have static real and positive permittivity $\epsilon_d$. In Fig.~\ref{figset}~{\bf (a)}, we give a specific example of the system being studied by choosing the radius of the nanoparticles as $R=25$nm and the distance between nanoparticles in the array as $d=75$nm (however, the general model we will introduce allows arbitrary values to be chosen for these parameters and for all other physical parameters). We consider the metal nanoparticles in the array support electron charge density oscillations in the longitudinal (L) and transverse (T) directions, as shown in the inset of Fig.~\ref{figset}~{\bf (a)} and neglect multipolar interactions~\cite{point}. In the system depicted in the main part of the figure, due to the direction in which the electron charge density oscillates in the nanowires, which act as the surface plasmon source and drain on the left and right hand sides, the nanotips at the ends are oriented to excite/collect charge density oscillations in the L direction. For excitation and collection in the T direction, both nanowires should be rotated by 90 degrees either clockwise or anticlockwise. Further details regarding the nanowire orientations are discussed later.

We now introduce the Hamiltonian for the system, justifying the physical origin of each of the terms appearing. The total Hamiltonian describing the system in Fig.~\ref{figset}~{\bf (a)} is given by
\be
\hat{H}=\hat{H}_{np}+\hat{H}_{s}+\hat{H}_{d}+\hat{H}_{np,s}+\hat{H}_{np,d}.
\label{HamT}
\ee

Here, the first term describes the linear nanoparticle array system consisting of $n$ nanoparticles and is given by
\be
\hat{H}_{np}=\sum_{i=1}^{n}\hbar \omega_i \hat{a}_i^\dag \hat{a}_i+\sum_{[i,j]}\hbar g_{i,j}\left( \hat{a}^\dag_i \hat{a}_j+\hat{a}^\dag_j \hat{a}_i \right), 
\label{Hnp}
\ee
where $\omega_i$ is the natural frequency of the field oscillation at the $i$-th nanoparticle, $g_{i,j}$ is the coupling strength between the fields of the $i$-th and $j$-th nanoparticles,  $[i,j]$ denotes a summation over nearest neighbours $j$ for a given nanoparticle $i$, and the operators $\hat{a}^\dag_i$ ($\hat{a}_i$) represent the creation (annihilation) operators associated with a field excitation at nanoparticle site $i$ which obey bosonic commutation relations $[\hat{a}_i,\hat{a}^\dag_j]=\delta_{ij}$. Here, a macroscopic quantization of the fields is used, where the field modes are defined as localized solutions to Maxwell's equations satisfying the boundary conditions of the metal-dielectric interface~\cite{Waks}. In this case, the electron response is contained within the dielectric function of the metal~\cite{ER,TSPP}. We consider either $L$ or $T$ polarization along the array, suppressing the polarization index. In addition, while the model we investigate here is for a linear array of nanoparticles, the theory introduced can be applied to more complex arrangements of nanoparticles~\cite{YK,Kub,Baer}.

The first term in Eq.~(\ref{Hnp}) represents the free Hamiltonian of the fields at the nanoparticles, where $\omega_i$ satisfies the Fr\"ohlich criterion, ${\rm Re}[\epsilon_m(\omega_i)]=-2 \epsilon_d$~\cite{Brong,Maier}. This criterion considers the nanoparticles to be small enough compared to the operating wavelength such that only dipole-active excitations are important~\cite{Zayats}. Taking all nanoparticles to have the same permittivity $\epsilon_m$, the local frequencies can be set to be equal, $\omega_i=\omega_0,~\forall i$. Due to the spherical symmetry of the nanoparticles, these local frequencies are independent of the polarization. The second term in Eq.~(\ref{Hnp}) represents a nearest-neighbour coupling between the near-field at each nanoparticle. In order to justify the physical mechanism of this second term, we briefly provide the correspondence of the quantum description of the nanoparticle array to the classical description~\cite{Brong}. 

Consider a quantum state $\ket{\psi}=\prod_{i}\ket{\alpha_i}$, where $\ket{\alpha_i}= e^{-\frac{1}{2}|\alpha_i|^2}e^{\alpha_i \hat{a}^\dag_i}\ket{0}$ is a coherent state and $\alpha_i$ is the mean field amplitude at the $i$-th nanoparticle. Here, the electric field variation of a coherent state $\ket{\alpha}$ approaches that of the classical wave picture in the limit of large amplitude $\alpha$~\cite{Loudon}. Taking $\hat{H}_{np}$ and $\ket{\psi}$ and substituting them into the Schr\"odinger equation, $i \hbar \frac{\partial}{\partial t}\ket{\psi}=\hat{H}_{np}\ket{\psi}$, one finds the differential equation for the mean field amplitudes as~\cite{YK}
\be
\frac{d \alpha_i}{d t}=-i\omega_0 \alpha_i-i\sum_{[i,j]}g_{i,j}\alpha_j.
\label{diff}
\ee
By choosing all the couplings to be equal $g_{i,j}=g=\frac{1}{2} \frac{\omega_I^2}{\omega_0}\gamma$, where $\gamma = \gamma_T=1$ and $\gamma = \gamma_L=-2$ are the relative couplings and phases for polarization $T$ and $L$ respectively (at a fixed distance $d$, array orientation and nanoparticle size $R$~\cite{Brong}), the differential equation in Eq.~(\ref{diff}) is exactly the same as the classical differential equation for the amplitude of the dipole moment $p_i$ (associated with the electric field at site $i$) for an array of interacting Hertzian dipoles under the condition $\omega_I \ll \omega_0$ for the interaction frequency $\omega_I$~\cite{Brong,YK}. This is a {\it weak coupling} approximation including only the nearest-neighbour interactions. In the classical Hertzian model, the dominant interaction in the system is considered to be between the nanoparticle dipoles via the F\"orster field, which has a $1/d^3$ dependence for $d \ll \lambda$, where $\lambda=\lambda_0/\sqrt{\epsilon_d}$ and $\lambda_0$ is the free-space wavelength corresponding to the natural frequency $\omega_0$ of the nanoparticle dipole field, $\lambda_0=2 \pi c/\omega_0$ ($c$ is the velocity of light in a vacuum)~\cite{Brong,Greiner}. This regime ($d \ll \lambda$) is known as the {\it near-field} approximation. Furthermore, the dipoles are considered point-like for $R \lesssim d/3$~\cite{point}, known as the {\it point-dipole} approximation. Thus, under the {\it weak coupling}, {\it near-field}, and {\it point-dipole} approximations, the quantum model with $g=\omega_I^2 \gamma/2\omega_0$ recovers the classical dynamics in the correct limit using coherent states. Here, the interaction frequency is given by $\omega_I=[e^2 \rho_{\rm el}R^3/3m^* \epsilon_0\epsilon_d d^3]^{1/2}$~\cite{Brong}, where $e$ is the electronic charge, $\rho_{\rm el}$ is the free electron density of the metal, $m^*$ is the optical effective electron mass and $\epsilon_0$ is the free-space permittivity. 

In Fig.~\ref{figset}~{\bf (b)} we show an example of the dependence of the magnitude of the coupling $g=\omega_I^2 \gamma/2\omega_0$ (in units of $\omega_0$) as the ratio of $R/d$ increases. Here we have taken the permittivity of the metal $\epsilon_m(\omega)$ as silver and used $\epsilon_m(\omega)=\epsilon_{\infty}-\omega_p^2/(\omega^2+i \omega \Gamma)+i/2$, where $\epsilon_{\infty}=5$, $\Gamma=6.25\times 10^{13}$ rad/s and $\omega_p=1.402 \times 10^{16}$ rad/s, which are chosen to obtain a best fit to experimental data at frequencies corresponding to freespace wavelengths $\lambda_0\gtrsim350$nm~\cite{Palik}, {\it i.e.} the optical range and above. This leads to $\omega_0=5 \times 10^{15}$ rad/s, the local frequency of the nanoparticles. In addition, we have used $\rho_{\rm el}=5.85 \times 10^{28}$m$^{-3}$, $m^*=8.7 \times 10^{-31}$kg and $\epsilon_d=1$~\cite{Brong}. The weak coupling approximation is equivalent to $|g| \ll \omega_0$ and we impose this by setting ${\rm max}|g|=0.1 \omega_0$. Note from Fig.~\ref{figset}~{\bf (b)} that the condition $\mathrm{max} |g| = 0.1 \omega_0$ satisfies the point-dipole approximation immediately, as well as the weak coupling for both polarizations. For the near-field approximation to also be satisfied we require $d \ll 2 \pi c /\omega_0\simeq377$nm. The example in Fig.~\ref{figset}~{\bf (a)} with $d=75$nm and $R=25$nm with silver satisfies all three of the required approximations. 

An additional requirement for the system is that quantum effects other than those due to the quantized surface plasmon field, such as electron tunneling between nanoparticles and the quantum size effect of each nanoparticle~\cite{Kreibig}, are negligible. This puts a lower limit on the distance $d$ between nanoparticles at~$\sim$~$1$nm~\cite{quantnp}, and nanoparticle radii of the order of $1$nm~\cite{Maier}, respectively. However, in order to confidently use the macroscopic approach for the quantization of the surface plasmon field due to the electron response, we assume nanoparticle radii $R\gtrsim 10$nm and therefore $d\gtrsim 30$nm to satisfy the point-dipole approximation. As far as we are aware it is still an open question as to what dimension the macroscopic approach to surface plasmon quantization breaks down. In addition, for the moment, we also neglect internal electronic relaxation at the nanoparticles and relaxation of the dipoles into the far-field. Damping will be introduced after the ideal case has been developed in the next section.

Continuing with our description of the physical system, the second and third terms of Eq.~(\ref{HamT}) represent the free Hamiltonian of surface plasmon fields in the source and drain nanowires on the left and right hand sides of Fig.~\ref{figset}~{\bf (a)} respectively and are given by
\bqa
{\hat H}_{s}&=&\int_{-\infty}^{\infty} d\omega \hbar \omega \hat{s}^\dag(\omega) \hat{s}(\omega), \nonumber\\
{\hat H}_{d}&=&\int_{-\infty}^{\infty} d\omega \hbar \omega \hat{d}^\dag(\omega) \hat{d}(\omega).\nonumber
\eqa
The operators of the nanowires correspond to continuum modes of surface plasmons, which obey the bosonic commutation relations $[\hat{s}(\omega),\hat{s}^\dag(\omega')]=\delta (\omega-\omega')$ and $[\hat{d}(\omega),\hat{d}^\dag(\omega')]=\delta (\omega-\omega')$. Again, a macroscopic quantization is carried out for the field~\cite{Chang3,TSPP,ER} and we have extended the integration of $\omega$ to cover the range $-\infty$ to $\infty$~\cite{Loudon}.

The surface plasmon excitation in each nanowire is taken to correspond to the fundamental transverse magnetic mode with winding number $m=0$~\cite{Chang3}. It can be generated by various methods. For instance, it could be generated via coupling of a photon from the far-field by focusing the quantized light field onto a grating structure at a thicker part of the tapered nanowire~\cite{grating}. Another method could be to use end-fire coupling of photons in conventional silica waveguides to the metal nanowires, again at a much thicker part of the tapered wire~\cite{Chang4}. One could also generate the plasmon excitations directly on the wires very close to the tip region by driving emitter systems, such as quantum dots~\cite{Lukin1} or NV-centers~\cite{Kol}, with coherent light to further reduce losses during propagation of the input~\cite{Chang3}. Here, the combination of metal and emitter system may provide additional flexibility in optimizing the field profile that couples to the nanoparticle - similar to a nanoantenna system~\cite{Agio} - rather than a direct coupling of the emitter system on its own.

The fourth and fifth terms of Eq.~(\ref{HamT}) represent the coupling of the surface plasmon field of the source nanowire to the LSP field of nanoparticle $1$ and the surface plasmon field of the drain nanowire to the LSP field of nanoparticle $n$ respectively. Using a weak-field linearized model~\cite{CG,WM} the terms are given by
\bqa
\hat{H}_{np,s}&=& i \hbar \int_{-\infty}^{\infty}  d \omega g_{\rm in}(\omega)[\hat{a}_1\hat{s}^\dag(\omega)-\hat{s}(\omega)\hat{a}_1^\dag], \nonumber\\
\hat{H}_{np,d}&=& i \hbar \int_{-\infty}^{\infty}  d \omega g_{\rm out}(\omega)[\hat{a}_n\hat{d}^\dag(\omega)-\hat{d}(\omega)\hat{a}_n^\dag].\nonumber
\eqa
Here the coupling parameters $g_{\rm in/out}(\omega)$ depend on the strength of the near-field coupling between the nanowires and nanoparticles. Focusing on the case of the source nanowire-to-nanoparticle coupling and considering a propagating surface plasmon in the nanowire entering the region at the nanotip from the left hand side, we have that for an appropriate paraboloidal profile of the nanowire, the mode function of the excitation near the tip strongly couples to a dipole orientated in the direction of the propagation and placed in close proximity~\cite{Chang3,Issa}. Thus, for the orientation of the source nanotip shown in Fig.~\ref{figset}~{\bf (a)}, the nanowire field couples predominantly to the $L$ polarized oscillation in the nanoparticle. For coupling to the transverse polarization, we rotate the nanotip clockwise by 90 degrees. The reciprocal case holds at the drain nanotip and in the orientation shown in Fig.~\ref{figset}~{\bf (a)}, the drain predominantly couples to $L$ polarized field oscillations in the $n$-th nanoparticle.

\begin{figure}[t]
\centering
\includegraphics[width=7.5cm]{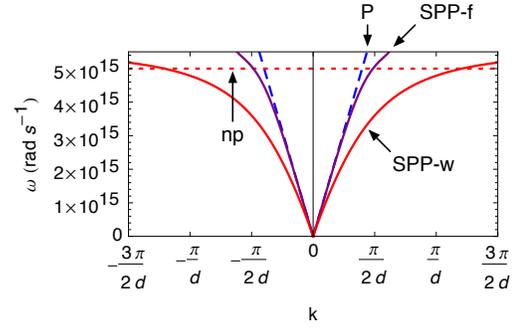}
\caption{Dispersion relation for various forms of surface plasmon excitation. Here silver has been chosen as the metal and air as the dielectric background media. The straight dashed line {\sf P} shows the photon dispersion relation in free-space, the curve {\sf SPP-f} is the dispersion relation of a surface plasmon field at a standard metal-air interface and the curve {\sf SPP-w} is the dispersion relation of a surface plasmon field in the tapered metal waveguide in the region of its tip at a radius of $R=25$nm. The horizontal dotted line {\sf np} is the natural frequency of a single nanoparticle $\omega_0$. For all curves only the real part of the wavenumber $k$ is plotted. The imaginary part being several orders of magnitude smaller.}
\label{figDR}
\end{figure}

Regardless of the excitation method of the plasmons in the nanowire, for a wire that has a slowly varying radius $R(x)$ with distance $x$ from the tip along the wire, we have a locally varying dispersion relation given by~\cite{Stock}
\be
\frac{\epsilon_m}{\kappa_m} \frac{I_1(k_0 \kappa_m R(x))}{I_0(k_0 \kappa_m R(x))}+\frac{\epsilon_d}{\kappa_d} \frac{K_1(k_0 \kappa_d R(x))}{K_0(k_0 \kappa_d R(x))} =0,
\label{wireDR}
\ee
where $I_p$ and $K_p$ are the modified Bessel functions, $\kappa_m=\sqrt{n^2-\epsilon_m}$, $\kappa_d=\sqrt{n^2-\epsilon_d}$, $k_0=\omega/c$ is the freespace wavenumber at a given frequency $\omega$ and $n=n(x,\omega)$ is the local effective refractive index at position $x$ along the nanowire. The radius of the nanotip at the end of the source wire defines the effective radius of the wire in the region just before the tip. We can use this to determine the approximate dispersion relation of the surface plasmons entering the nanotip region where they couple to the LSP of the first nanoparticle. Therefore, Eq.~(\ref{wireDR}) can be solved for a given set of physical parameters in order to obtain the local effective refractive index $n(x,\omega)$, leading to the dispersion relation $k=n(x,\omega)\omega/c$ for the surface plasmons close to the tip ($x\simeq0$). In Fig.~\ref{figDR} we show the dispersion relation for a free-space photon $k=k_0=\omega/c$, a nanowire surface plasmon (for $R(0)=25$nm) using Eq.~(\ref{wireDR}), a standard metal-air interface surface plasmon $k=(\omega/c)\sqrt{\epsilon_m(\omega)/(1+\epsilon_m(\omega))}$~\cite{TSPP} and the nanoparticle natural oscillation frequency $\omega_0$. In all cases the example metal is taken to be silver with the dielectric function defined previously.  Here we have chosen to represent the wavenumber $k$ in units of an array spacing $d=75$nm. Note that only the surface plasmon field from the tip region of a nanowire has the potential to achieve \textit{both} the correct energy conservation ($\omega$ matching) and dipole-coupling~\cite{Chang3,Issa} for efficient near-field coupling to the first (or last) nanoparticle of the array. 

Thus, by setting the coupling $g_{\rm in}(\omega)$ in $\hat{H}_{np,s}$ according to the physical geometries and tip orientation being considered, one can model coupling of the surface plasmon in the source nanowire to the first nanoparticle and its reflection back along the nanowire. Similarly, by setting the coupling $g_{\rm out}(\omega)$ in $\hat{H}_{np,d}$, one can model coupling of the last nanoparticle's near-field to the drain nanowire and its reflection back along the nanoparticle array. As the field profiles at the tips are similar in form to those of the nanoparticles, the same physical approximations as the inter-particle coupling strengths $g_{i,j}$ should be satisfied by the couplings $g_{\rm in}(\omega)$ and $g_{\rm out}(\omega)$ in order for the model to be a consistent description. The $g_{\rm in/out}(\omega)$ couplings can then be modified to model non-ideal mode function profiles due to the tip shape and other geometrical factors. 
 
\subsection{Transmission and dispersion}

We now use the Hamiltonian in Eq.~(\ref{HamT}) to model the transmission of a quantum state injected into the array by the source nanowire and then its propagation along the array until it is subsequently extracted out by the drain nanowire. In order to do this we use an effective scattering matrix approach that will link the input field operators of the source nanowire to the output field operators of the drain nanowire, providing a method to map arbitrary input quantum states to output quantum states. This scattering matrix is obtained by applying input-output formalism~\cite{CG,WM} to the nanoparticle array, as summarized in Appendix A. The benefit of this approach is that we may treat the nanoparticle array as a waveguide with an effective medium, which makes the description of the system in the context of the transfer of quantum states more intuitive. However, it is important to note that one can also use this approach to investigate the internal quantum dynamics of the nanoparticle array and even interactions with other resonant systems. For instance, emitter systems such as NV centres, placed in close proximity~\cite{CG,WM}. 

Using the input-output formalism in Appendix A, we have the relation between input field operators $\hat{s}_{\rm in}(\omega)$ and $\hat{d}_{\rm in}(\omega)$, and output field operators $\hat{s}_{\rm out}(\omega)$ and $\hat{d}_{\rm out}(\omega)$ for the nanowires as follows,
\bqa
\hat{s}_{\rm in}(\omega)&=& {\cal R}_s^*(\omega)\hat{s}_{\rm out}(\omega)+{\cal T}^*_s(\omega)\hat{d}_{\rm out}(\omega), \label{inop1} \\
\hat{d}_{\rm in}(\omega)&=& {\cal T}^*_d(\omega)\hat{s}_{\rm out}(\omega)+{\cal R}^*_d(\omega)\hat{d}_{\rm out}(\omega), \label{inop2}
\eqa 
where the transmission ${\cal T}_{s,d}$ and reflection ${\cal R}_{s,d}$ coefficients are functions of the system parameters $g_{\rm in}$, $g_{\rm out}$, $g_{i,j}$ and $\omega_i$, and the relation $|{\cal R}_{s,d}(\omega)|^2+|{\cal T}_{s,d}(\omega)|^2=1$. Taking the Hermitian conjugate of Eq.~(\ref{inop1}) we have $\hat{s}_{\rm in}^\dag(\omega)={\cal R}_s(\omega)\hat{s}_{\rm out}^\dag(\omega)+{\cal T}_s(\omega)\hat{d}_{\rm out}^\dag(\omega)$. This allows us to describe the nanoparticle array as an effective waveguide, with transmission ${\cal T}_s(\omega) = |{\cal T}_s(\omega)|e^{i(kx\pm\pi)}$, where $x=(n+1)d$ is the total effective distance (from the centre of the source tip to the centre of the drain tip, as shown in Fig.~\ref{figset}~(a)). The factor $\pm\pi$ takes into account the phase difference of $-1$ in the definition between the input and output field operators and the wavenumber $k$ depends on the system parameters $g_{\rm in}$, $g_{\rm out}$, $g_{i,j}$ and $\omega_i$. We now drop the index $s$ in the transmission coefficient for ease of notation and consider only transmission in the forward direction. Thus, with the use of ${\cal T}(\omega)$, we obtain
\be
k= \frac{{\rm arg}\left[{\cal T}(\omega)\right]\mp \pi + 2 m\pi}{(n+1)d}, m=0,\pm1,\pm2,\dots, 
\label{knp}
\ee
where the additional factor of $2m\pi$ is included to reflect the cyclical degeneracy of the wavenumber. 
\begin{figure}[t]
\centering
\includegraphics[width=8.5cm]{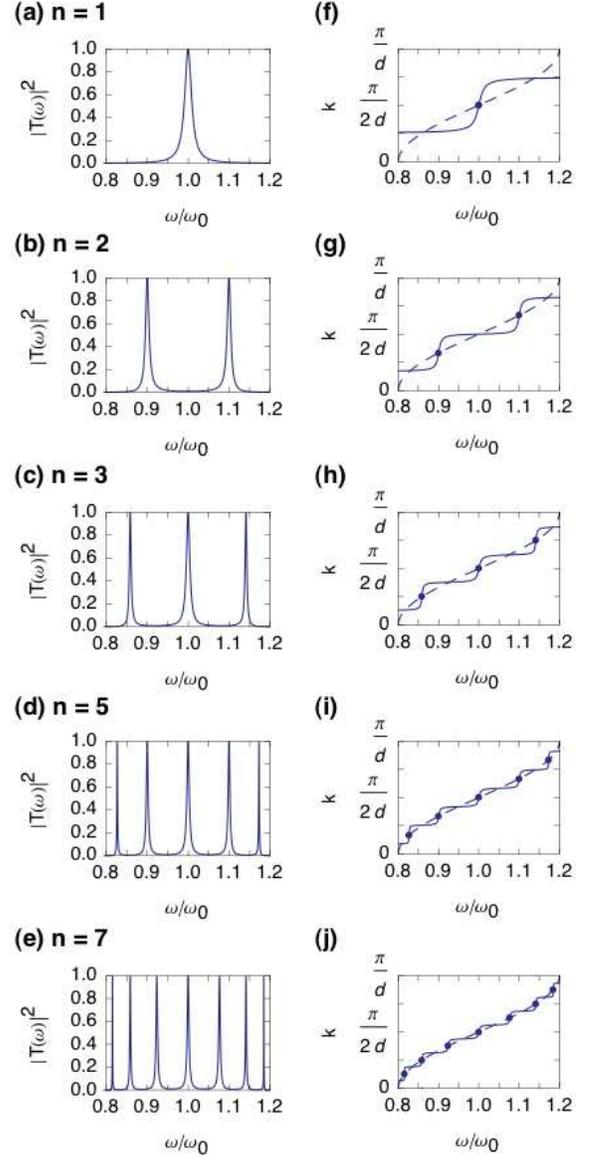}
\caption{Transmission spectral profiles and dispersion relations for undamped arrays of nanoparticles. Panels {\bf (a)-(e)} correspond to the amplitude squared of the transmission, $|{\cal T}(\omega)|^2$, as the frequency is varied for an array of $n=1,~2,~3,~5$ and 7 nanoparticles respectively. Panels {\bf (f)-(j)} correspond to plots of the effective wavenumber $k$ for $n=1,~2,~3,~5$ and 7 respectively. Also shown are points corresponding to the $k^j$ transmission resonance peaks from panels {\bf (a)-(e)} as well as the dispersion relation for the infinite array case (dotted line).}
\label{figresonance}
\end{figure}

In Fig.~\ref{figresonance}~(a)-(e) we plot the amplitude squared of the transmission, $|{\cal T}(\omega)|^2$, as the frequency is varied for an array of $n=1,~2,~3,~5$ and 7 nanoparticles respectively. While an analytical form for ${\cal T}(\omega)$ can be found, due to the general complexity of all the system parameters, here we show only explicit examples where we have taken all local frequencies to be equal $\omega_i=\omega_0,\forall i$ and the couplings to be equal $g_{i,j}=g_{\rm np}=-0.1 \omega_0, \forall i$ and its nearest neighbors $j$ (with the minus sign for the longitudinal polarization, as $\gamma_L=-2$). In Appendix B we provide the analytical form for the ${\cal T}(\omega)$'s. Physically, this chosen coupling regime corresponds to an array with $d/R\simeq3$, for example $R=25$nm and $d=75$nm, if we take the metal to be silver as before. The source and drain couplings are set as $g_{\rm in/out}=0.01\omega_0$, achieved by varying the distance between the nanowire tips and their respective nearest nanoparticle. For a given number of nanoparticles $n$, the transmission spectral profiles in Fig.~\ref{figresonance}~(a)-(e) have $n$ resonances at frequencies $\omega_{r_j}=\omega_0+2g_{\rm np} \cos (k^j d)$, where $k^j=j\pi/(n+1)d$ for $j=1,\dots,n$. In Fig.~\ref{figresonance}~(f)-(j) we plot the effective wavenumber $k$ from Eq.~(\ref{knp}) for $n=1,~2,~3,~5$ and 7 respectively. Points corresponding to the $k^j$ transmission resonance peaks from Fig.~\ref{figresonance}~(a)-(e) are marked as circles. Also included in these figures is the dispersion relation for the infinite array case (dashed line), where the $k^j$ take on continuous values~\cite{YK}, with a positive group velocity over the entire range due to taking the minus sign (phase) for the longitudinal polarization coupling $g$ in this example. From Fig.~\ref{figresonance}~(f)-(j) one can clearly see that as $n$ is increased, the band structure of the infinite array case is gradually recovered, where each $(\omega_{r_j},k^j)$ point corresponds to the dominant excitation of a stationary eigenstate of the system Hamiltonian $\hat{H}_{np}$, analogous to the case of coupled cavities~\cite{Fran}, for instance in the case of photonic crystals~\cite{Notomi}. 
\begin{figure*}[t]
\centering
\includegraphics[width=16cm]{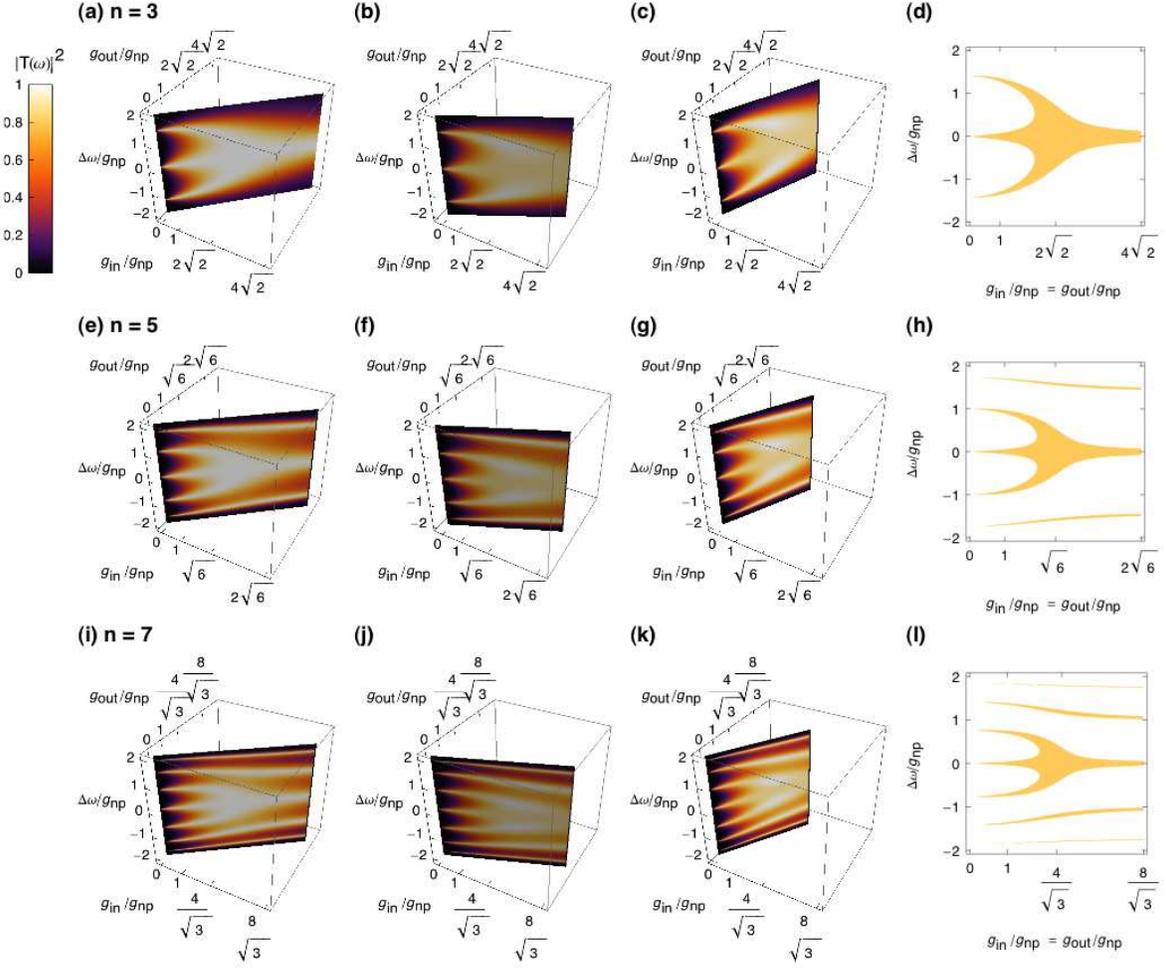}
\caption{Cross-sections of the amplitude squared of the transmission, $|{\cal T}(\omega)|^2$, as the frequency $\omega$ and couplings $g_{\rm in}$ and $g_{\rm out}$ are varied. Here, $\Delta \omega = \omega-\omega_0$, where $\omega_0$ is the resonant frequency of each nanoparticle. Panels {\bf (a)},~{\bf (e)} and {\bf (i)} correspond to couplings $g_{\rm in}=g_{\rm out}$ for an array of $n=3,~5$ and 7 nanoparticles respectively. Panels {\bf (b)},~{\bf (f)} and {\bf (j)} correspond to $g_{\rm in}=2g_{\rm out}$ and panels {\bf (c)},~{\bf (g)} and {\bf (k)} correspond to $g_{\rm in}=g_{\rm out}/2$. Panels {\bf (d)},~{\bf (h)} and {\bf (l)} have a threshold placed on the value of $|{\cal T}(\omega)|^2$, with the solid area corresponding to $|{\cal T}(\omega)|^2 \geq 0.98$ for $g_{\rm in}=g_{\rm out}$.}
\label{arbnodamp}
\end{figure*}

Note that while the above examples provide a basic insight into the system dynamics, the formalism introduced here can be used to describe more complex and general plasmonic nanoparticle systems with arbitrary couplings and natural local frequencies. 
We now proceed to focus on odd numbered nanoparticle systems with $n>1$ in order to understand the transmission properties of larger arrays in more general regimes. A similar study could be made for even numbered systems, however, we choose odd numbered as there is always a resonance at the natural frequency $\omega_0$. This will become important later in our study of quantum state transfer. 

In Fig.~\ref{arbnodamp}~(a),~(e) and (i) we plot a cross-section of the amplitude squared of the transmission, $|{\cal T}(\omega)|^2$, as the frequency $\omega$ and coupling $g_{\rm in}~(=g_{\rm out})$ is varied for an array of $n=3,~5$ and 7 nanoparticles respectively. Here, $\Delta \omega = \omega-\omega_0$ and we have chosen to plot all parameters in units of the nanoparticle coupling $g_{i,j}=g_{\rm np},~\forall i$ and its nearest neighbors $j$. The plots are therefore independent of $g_{\rm np}$, as long as $g_{\rm np} \ll \omega_0$ is satisfied. Increasing (decreasing) $g_{\rm np}$ shrinks (expands) all axes. This observation can be useful when comparing two $g_{\rm np}$ regimes with each other. Note also that ${\rm max}(g_{\rm in}/g_{\rm np}^{\rm max},g_{\rm out}/g_{\rm np}^{\rm max})=1$ must be imposed, where $g_{\rm np}^{\rm max}=0.1 \omega_0$, otherwise we would move away from the weak coupling regime for the source and drain. In other words, the rescaled couplings $g_{\rm in}/g_{\rm np}$ and $g_{\rm out}/g_{\rm np}$ can in principle go higher than 1, but the value for $g_{\rm np}$ must be lower than $0.1 \omega_0$ to compensate so that we are still in the weak coupling regime. In Fig.~\ref{arbnodamp}~(b),~(f) and (j) we plot a different cross-section for $n=3,~5$ and 7 nanoparticles, where $g_{\rm in}=2g_{\rm out}$ and in Fig.~\ref{arbnodamp}~(c),~(g) and (k) for $g_{\rm in}=g_{\rm out}/2$. In Fig.~\ref{arbnodamp}~(d),~(h) and (l) we place a threshold on the value of $|{\cal T}(\omega)|^2$ such that the solid red area corresponds to $|{\cal T}(\omega)|^2 \geq 0.98$ for $g_{\rm in}=g_{\rm out}$. One can see that as the source and drain couplings increase, the range over which the transmission is close to ideal becomes enlarged about the central resonance, although if the couplings are too large this range reduces back again. Similar behaviour can be seen for larger odd numbers of nanoparticles, with the central `fork' area becoming narrower as $n$ increases. These behaviors can be understood as follows. The early increase of $g_\mathrm{in}$ enables the off-resonant transfer from the source to the first nanoparticle, whereas its late increase leads to strong coupling as if the first nanoparticle becomes the extended `tip' of the nanotip. A similar argument about $g_\mathrm{out}$ applies for the last nanoparticle and the drain nanotip. Thus the large $g_\mathrm{in/out}$ implies that the number of nanoparticles is effectively reduced to $n-2$. In the moderate magnitude of $g_\mathrm{in/out}$, we have the broad region of frequency $\omega$ for highly efficient transfer. This observation will be important in our study of quantum state transfer in the next section. 

\section{Quantum state transfer}
\subsection{Qubit transfer}

We now consider quantum information in the form of a single quantum bit, or qubit, transferred across a metal nanoparticle array. We write the input qubit state in the source as $\ket{\psi}_s=a \ket{0}_s+b \ket{1_\xi}_s$~\cite{Pegg,Resch}, where $|a|^2+|b|^2=1$, and $\ket{0}_s$ and $\ket{1_\xi}_s$ represent the vacuum state and single plasmon wavepacket in the source (at the tip), respectively. The wavepacket is characterized by a spectral profile $\xi(\omega)$ with $\int_{-\infty}^{\infty} {\rm d} \omega |\xi(\omega)|^{2}=1$. More explicitly we have
\be
\ket{\psi}_s=a\ket{0}_s+b \int_{-\infty}^{\infty} {\rm d}\omega \xi(\omega)\hat{s}^\dag_{\rm in}(\omega)\ket{0}_s.
\label{instat}
\ee
Then for a given input state from the source, we take both the nanoparticles and drain to be initially in the vacuum state. Using the relation in Eq.~(\ref{inop1}) and substituting for $\hat{s}^\dag_{\rm in}(\omega)$, then tracing out the state in the source (see Appendix C), we obtain the output state in the drain nanowire (at the tip) as 
\bqa
\rho_d&=&\bigg(|a|^2+|b|^2\int_{-\infty}^{\infty} {\rm d}\omega |\xi(\omega)|^2(1-|{\cal T}(\omega)|^2)\bigg) \pro{0}{d}{0} \nonumber \\
&& + a b^*\int_{-\infty}^{\infty} {\rm d}\omega \xi^*(\omega){\cal T}^*(\omega) \pro{0}{d}{1_{\omega}} \nonumber \\
&& +a^* b \int_{-\infty}^{\infty} {\rm d}\omega \xi(\omega){\cal T}(\omega) \pro{1_{\omega}}{d}{0} \nonumber \\
&&+|b|^2\int_{-\infty}^{\infty} {\rm d}\omega \int_{-\infty}^{\infty} {\rm d}\omega'\xi(\omega)\xi^*(\omega') \times \nonumber \\
&& \qquad \qquad \qquad \qquad ~~~{\cal T}(\omega){\cal T}^*(\omega')\pro{1_{\omega}}{d}{1_{\omega'}}, \label{outstat}
\eqa
where $\ket{1_{\omega}}_d=\hat{d}^\dag_{\rm out}(\omega)\ket{0}_d$. For perfect state transfer, {\it i.e.} ${\cal T}(\omega)=e^{i(kx\pm\pi)}$, giving $|{\cal T}(\omega)|^2=1$ and $|{\cal R}(\omega)|^2=0$, the output state going into the drain nanowire, described by Eq.~(\ref{instat}), becomes the pure state $\rho_d=\pro{\psi'}{d}{\psi'}$, where $\ket{\psi'}_d$ is equivalent to Eq.~(\ref{instat}), but with $\xi(\omega) \to \xi'(\omega)=\xi(\omega)e^{i(kx\pm\pi)}$ and $\hat{s}_{\rm in}^\dag(\omega) \to \hat{d}_{\rm out}^\dag(\omega)$. The change in the spectral amplitude of the wavepacket is equivalent (upon Fourier transform) to a positive temporal shift (delay) in the wavepacket, which corresponds to the time that the wavepacket takes to move from the source tip to the drain tip. For concreteness, consider a Gaussian wavepacket with spectral amplitude profile
\be
\xi(\omega)=(2\pi\sigma^2)^{-1/4}e^{-\frac{(\omega_0-\omega)^2}{4 \sigma^2}}\label{gaussian},
\ee
where $\omega_0$ is the central frequency and $\sigma=\delta \omega/(2\sqrt{2 {\rm ln}2})$ is the standard deviation corresponding to a FWHM bandwidth $\delta \omega$ for the spectral intensity profile $|\xi(\omega)|^2$. Applying the transform $\xi(\omega) \to \xi'(\omega)=\xi(\omega)e^{i(kx\pm\pi)}$ and assuming a small enough $\delta \omega$ so that there is linear dispersion about $\omega_0$, then $k\simeq\omega n_\mathrm{eff}/c=\omega/c_\mathrm{eff}$, where $n_\mathrm{eff}$ and $c_\mathrm{eff}$ are the effective refractive index and speed across the nanoparticle array. We can then write $kx \simeq \omega x/c_\mathrm{eff}=\omega \delta t$, where $\delta t$ is the time taken for the wavepacket to propagate from the source tip to the drain tip and $x$ is the total effective distance (from the centre of the source tip to the centre of the drain tip). Setting $\xi(\omega) \to \xi'(\omega)=\pm \xi(\omega)e^{i\omega \delta t}$ and taking the Fourier transform one finds
\be\label{shiftedgaussian}
\xi'(t)=\pm(2\sigma^2/\pi)^{1/4}e^{-\sigma^2(t-\delta t)^2 -i \omega_0(t-\delta t)} \equiv \xi(t-\delta t),
\ee
corresponding to a positive shift, or delay, of $\delta t$ in the time domain.

We now consider the fidelity of the transfer, defined as $F=~_{d}\bra{\psi'}\rho_d\ket{\psi'}_{d}$, where $\ket{\psi'}_{d}$ is the ideal transferred state including the dispersion, as defined previously. The fidelity describes how close the output state is to the expected one, being zero for orthogonal states and 1 for perfect transfer. Thus we use it to quantify the quality of state transfer. A straightforward substitution gives the more explicit form
\bqa \label{fidout}
F=|a|^4 &+& |a|^2|b|^2\int_{-\infty}^{\infty} {\rm d}\omega |\xi(\omega)|^2(1-|{\cal T}(\omega)|^2+2|{\cal T}(\omega)|) \nonumber \\
&&+|b|^4 \left( \int_{-\infty}^{\infty} {\rm d}\omega |\xi(\omega)|^{2} |{\cal T}(\omega)| \right)^{2} . 
\eqa
Using the Bloch sphere coordinates $a=\cos (\theta/2)$ and $b=e^{i \phi} \sin (\theta/2)$ and averaging the fidelity over all possible qubit states $\bar{F}=\frac{1}{4\pi}\int_0^{\pi}{\rm d} \theta\int_{0}^{2\pi}{\rm d} \phi F \sin \theta$, one finds $|a|^4\to 1/3$, $|b|^4\to 1/3$ and $|a|^2|b|^2\to 1/6$. Thus, for a given nanoparticle array and input wavepacket defined by $\xi(\omega)$, with a knowledge of $|{\cal T}(\omega)|$, one can calculate the average fidelity of the output qubit state going into the drain nanowire using Eq.~(\ref{fidout}). Note that Eq.~(\ref{fidout}) is irrespective of dispersion and depends only on $|{\cal T}(\omega)|$, since we have taken the fidelity with respect to $\ket{\psi'}_{d}$, setting $\xi'(\omega)$ correctly to the expected profile resulting from an arbitrary input $\xi(\omega)$, which compensates the dispersion of transmission. However, for simplicity we limit our discussion to linear dispersion, where the expected output state by perfect transfer has the profile given in Eq.~(\ref{shiftedgaussian}).
\begin{figure*}[t]
\centering
\includegraphics[width=14cm]{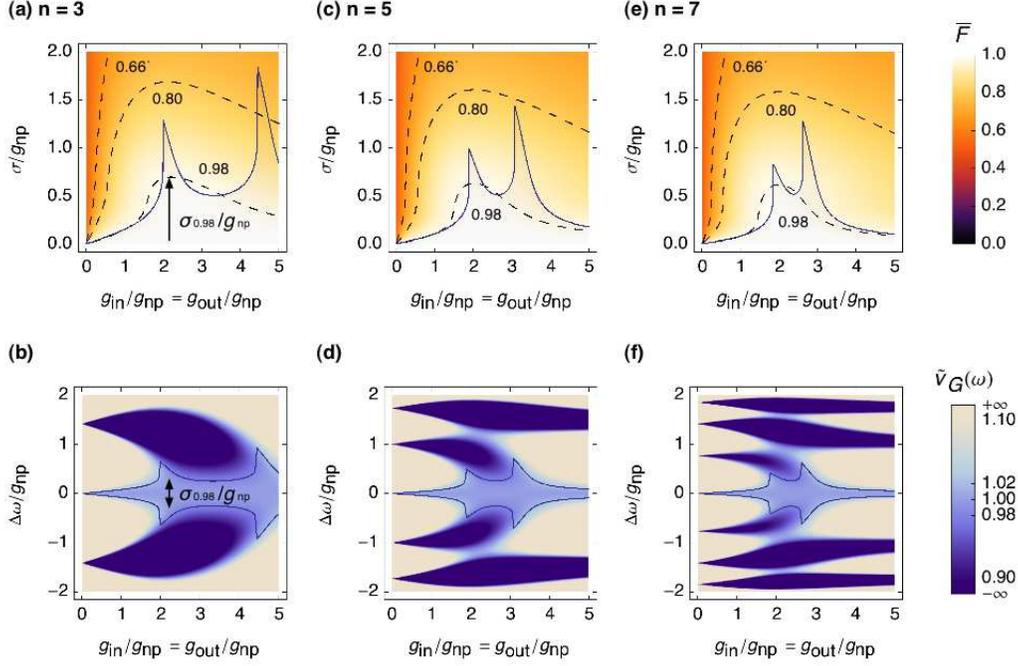}
\caption{Fidelity of quantum state transfer for a single qubit wavepacket in the absence of damping and scaled group velocity over a range of frequencies and couplings. {\bf (a):} Average fidelity $\bar{F}$ for transferring a qubit wavepacket over $n=3$ nanoparticles as the bandwidth ($\sigma$) and coupling $g_{\rm in}~(=g_{\rm out})$ are modified. All parameters are scaled by $g_{\rm np}$ {\bf (b):} Scaled group velocity $\tilde{v}_G(\omega)=v_G(\omega)/v_G(\omega_0)$ as it deviates from that at the resonance frequency $\omega_0$ for $n=3$, showing the regions of approximate linear dispersion. {\bf (c):} Average fidelity for transferring a qubit wavepacket over $n=5$ nanoparticles. {\bf (d):} Scaled group velocity $\tilde{v}_G(\omega)$ as it deviates from that at the resonance frequency $\omega_0$ for $n=5$. {\bf (e):} Average fidelity for transferring a qubit wavepacket over $n=7$ nanoparticles. {\bf (f):} Scaled group velocity $\tilde{v}_G(\omega)$ as it deviates from that at the resonance frequency $\omega_0$ for $n=7$. In all group velocity plots, the region inside the blue lines for $\Delta \omega$ corresponds to the frequency range below the blue lines for $\sigma$ shown in the average fidelity plots.}
\label{fidnoloss}
\end{figure*}

In Fig.~\ref{fidnoloss}~(a),~(c) and (e) we show the average fidelity $\bar{F}$ for an array of $n=3$, 5 and 7 nanoparticles. Here one can see immediately that for a small enough bandwidth, the state can be transferred across the array with perfect fidelity. The dashed lines correspond to fidelity contours, with the lowest curve ($0.66^\cdot$) corresponding to the classical threshold for a quantum channel: the best fidelity achievable by measuring an unknown qubit along a random direction and then sending the result through a classical channel using classical correlations~\cite{fidlimit}. The solid blue curves bound the region (from below) in which the dispersion is approximately linear, so that we can use the approximation $k\simeq\omega n_\mathrm{eff}/c=\omega/c_\mathrm{eff}$ to obtain the form of the expected output spectral profile $\xi'(\omega)$. This region is found by calculating the group velocity $v_G(\omega)$, where $v^{-1}_{G}(\omega)= \partial k/ \partial \omega' |_{\omega'=\omega}$ and $k$ is found from Eq.~(\ref{knp}). For linear dispersion about the resonant frequency we should have that $v_G(\omega)\simeq v_G(\omega_0)$. In Fig.~\ref{fidnoloss}~(b),~(d) and (f) we show the scaled group velocity $\tilde{v}_G(\omega)=v_G(\omega)/v_G(\omega_0)$ for an array of $n=3$, 5 and 7 nanoparticles. One can see that there is a wide frequency range available in the linear dispersive regime, given a large enough input/output coupling can be achieved.

\subsection{Single-photon and coherent state transfer}

We now discuss the transfer of two particular kinds of input state: single-photon states and very low-intensity classical light described by coherent states having an average photon number of $1$. These are typical quantum and classical states of light, respectively, and while they appear to be similar, they are in fact very different states altogether, with different measurable physical properties. On one hand, a single-photon state injected into the source nanowire can be described by $\ket{1_{\xi}}_{s} =\int_{-\infty}^{\infty} {\rm d}\omega \xi(\omega)\hat{s}^\dag_{\rm in}(\omega)\ket{0}_{s}$, with $\int_{-\infty}^{\infty} {\rm d}\omega |\xi(\omega)|^{2}=1$. On the other hand, a coherent state is described by $\ket{\{\alpha\}}_{s} = \mathrm{exp}(\hat{s}_{{\rm in}, \alpha}^{\dagger}-\hat{s}_{{\rm in}, \alpha}) \ket{0}_{s}$, where the wavepacket operators are $\hat{s}_{{\rm in}, \alpha}^{\dagger} = \int_{-\infty}^{\infty} {\rm d}\omega \alpha(\omega)\hat{s}^\dag_{\rm in}(\omega)$, with $\int_{-\infty}^{\infty} {\rm d}\omega |\alpha(\omega)|^{2}=\left< \hat{n} \right>$~\cite{Loudon}. Using the quantum theory we have developed to describe the nanoparticle array system, we now highlight a difference between single-photon states and coherent states (which are consistent with classical electromagnetic theory). The aim is to show the necessity of our quantum formalism in order to correctly predict measurable physical properties of the transfer process. 

First we consider that the average photon number of the injected coherent state is 1, \textit{i.e.}, $\left< \hat{n} \right>=\int_{-\infty}^{\infty} {\rm d}\omega |\alpha(\omega)|^{2}=1$, and the wavepacket amplitude $\alpha(\omega)$ is the same Gaussian form as $\xi(\omega)$. The scattering matrix given in Eq.~(\ref{inop1}) enables us to treat the nanopartice array as an effective beam splitter, and for a single-photon state and coherent state we obtain the following respective output states at the nanotips
\bqa
\ket{1_{\xi}}_{s} &\rightarrow& \int_{-\infty}^{\infty} {\rm d}\omega  \bigg( \xi(\omega) {\cal R}(\omega) \ket{1_{\omega}}_{s}+ \xi(\omega){\cal T}(\omega)\ket{1_{\omega}}_{d} \bigg),\nonumber \\
\ket{\{\alpha\}}_{s} &\rightarrow& \ket{\{\ \alpha {\cal R} \}}_{s} \otimes \ket{\{\ \alpha {\cal T} \}}_{d}.\nonumber
\eqa
It is clear from the above that each input state arriving at the source nanotip is transmitted and reflected in a different way: the single-photon state becomes an entangled state of transmitted and reflected single-plasmon states while the coherent state remains as a separable state of transmitted and reflected coherent states of plasmons. Nevertheless, the detection probabilities (mean excitation flux) at the drain are exactly the same as each other. This is calculated by finding the expectation value $\left< \hat{n}_{d_{\rm out}} \right>$, where $\hat{n}_{d_{\rm out}}=\int_{-\infty}^{\infty} {\rm d}\omega \hat{d}_{\rm out}^{\dagger}(\omega)\hat{d}_{\rm out}(\omega)$, and gives the same result for both input states
\be 
\left< \hat{n}_{d_{\rm out}} \right> = \int_{-\infty}^{\infty} {\rm d}\omega |\xi(\omega)|^{2} |{\cal T}(\omega)|^{2}=\int_{-\infty}^{\infty} {\rm d}\omega |\alpha(\omega)|^{2} |{\cal T}(\omega)|^{2},\nonumber
\ee
This implies that there is no difference in the energy transfer efficiency between single-photon states and coherent states when they are injected into the nanoparticle array. However, in quantum information processing, and in particular quantum communication, a more meaningful measure of the transfer success is not the energy efficiency, but how well the information content that is encoded into a physical state is preserved. This can be quantified by the fidelity between the transferred state and the ideal transferred state, as defined in the previous section and it is a measurable physical property of the transfer process; it can be measured by performing quantum state tomography~\cite{NC}. The fidelity for the transfer of a single-photon state is obtained by substituting $a=0$ and $b=1$ in Eq.~(\ref{fidout}). The fidelity for the continuous-mode coherent state transfer is summarized in Appendix D. The respective fidelities are as follows,
\bqa
&& \hskip-0.0cm \left[ \int_{-\infty}^{\infty} {\rm d}\omega |\xi(\omega)|^{2} |{\cal T}(\omega)| \right]^{2} ~~ {\rm and}\nonumber \\
&& \hskip-1.0cm \mathrm{exp} \left[- \int_{-\infty}^{\infty} {\rm d}\omega |\alpha(\omega)|^{2}( |{\cal T}(\omega)|-1)^{2} \right].  \nonumber
\eqa
It is clear that they are not the same. It is important to note that while the transfer of a single-photon state and coherent state are equivalent in the sense that the nanoparticle array transmits the same amount of their energy from the source to the drain nanowire, they are in fact different from the viewpoint of the transfer of information encoded within the states. This behaviour naturally carries over to the general case of qubits, where $a$ and $b$ are arbitrary, as in Eq.~(\ref{fidout}). It also applies when damping is introduced (see next section).
\begin{figure}[b]
\centering
\includegraphics[width=8.8cm]{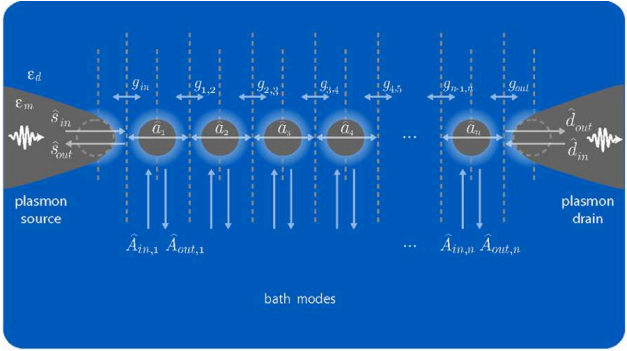}
\caption{The plasmonic nanoparticle array including bath modes to model damping at each nanoparticle.}
\label{dampchain}
\end{figure}

\section{Damping}
\subsection{Physical model and transmission}

\begin{figure}[t]
\centering
\includegraphics[width=8.0cm]{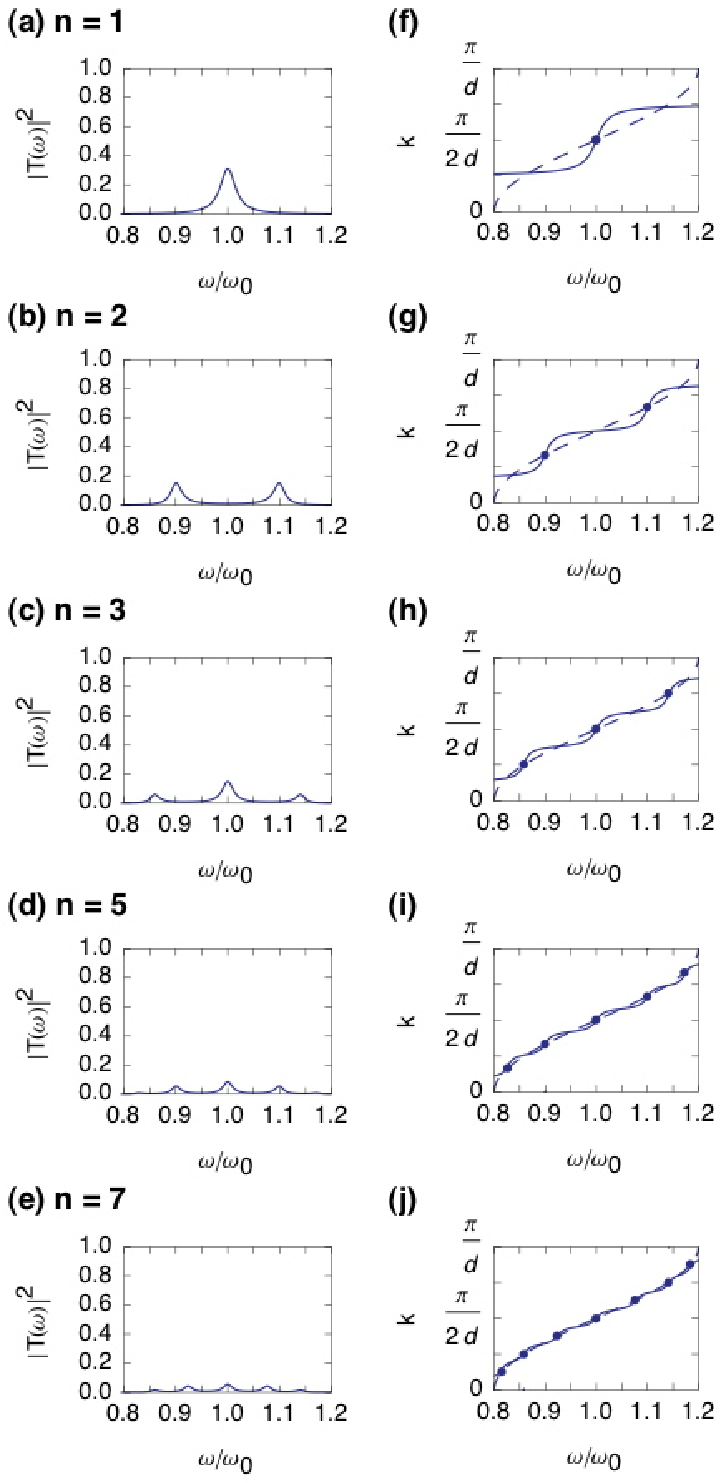}
\caption{Transmission spectral profiles and dispersion relations for damped arrays of nanoparticles. Panels {\bf (a)-(e)} correspond to the amplitude squared of the transmission, $|{\cal T}(\omega)|^2$, as the frequency is varied for an array of $n=1,~2,~3,~5$ and 7 nanoparticles respectively. Panels {\bf (f)-(j)} correspond to plots of the real part of the effective wavenumber $k^r$ for $n=1,~2,~3,~5$ and 7 respectively. Also shown are points corresponding to the $k^{r,j}$ transmission resonance peaks from panels {\bf (a)-(e)} as well as the dispersion relation for the infinite array case (dashed line). Here the couplings used are the same as the undamped case, {\it i.e.}, $g_{np}=-0.1\omega_{0}$ and $g_{\rm in/out}=0.01 \omega_0$. Note that larger transmission values can be achieved by increasing these couplings, as explained in the text and shown in Fig.~\ref{arbdamp}.}
\label{figresonancedamp}
\end{figure}
\begin{figure*}[t]
\centering
\includegraphics[width=14cm]{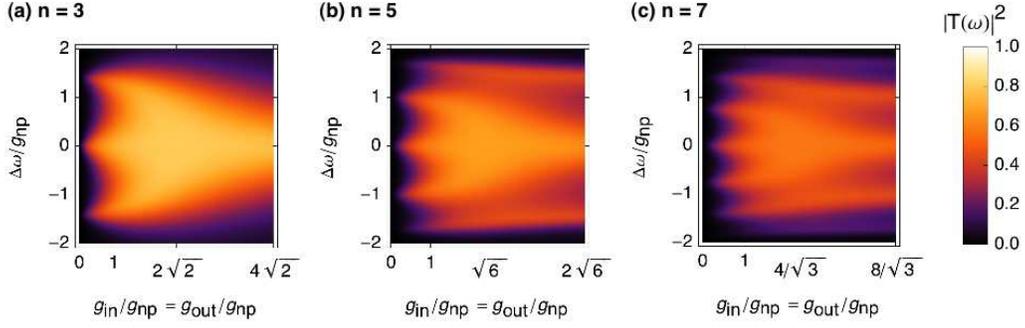}
\caption{Damping for the amplitude squared of the transmission, $|{\cal T}(\omega)|^2$, as the frequency $\omega$ and couplings $g_{\rm in}$ and $g_{\rm out}$ are varied. Panels {\bf (a)},~{\bf (b)} and {\bf (c)} correspond to couplings $g_{\rm in}=g_{\rm out}$ for an array of $n=3,~5$ and 7 nanoparticles respectively.}
\label{arbdamp}
\end{figure*}
We now include damping in our model. The effects of loss in the system are due to the interaction of the electrons (supporting the surface plasmon field) with phonons, lattice defects and impurities~\cite{JC,Brong}, as well as radiative scattering of the surface plasmon into the far-field~\cite{Brong}. For most scenarios of nanoparticle arrays, the couplings between nanoparticles are large enough such that most of the field remains within the array, with radiative scattering rates generally 5 orders of magnitude smaller than the relaxation rate~\cite{Brong}. Thus we assume radiative scattering can be neglected in our model. This assumption also allows us to neglect possible scattering at the tips. Electronic relaxation effects on the other hand cannot be neglected and lead to damping of the supported surface plasmon field. In our model we describe this as an amplitude damping channel at each nanoparticle. In this context a mechanism can be introduced where the damping is modeled by coupling of the field at each nanoparticle to an independent bath mode, which is eventually traced out from the system dynamics, as shown in Fig.~\ref{dampchain}. As we are interested in the mapping of the input field at the source tip to the output field at the drain tip, we assume that the source and drain excitations experience no loss when propagating in/out of the tip regions. Such insertion loss can however be incorporated using standard waveguide methods~\cite{Loudon,TSPP}, although a specific model will depend on how the fields in the nanowires are excited and collected, for instance, how far they propagate in the nanowires. Various types of dielectric-metal structures can significantly reduce these losses~\cite{tipdiel}. 

The scattering matrix in the presence of damping is derived in Appendix E. In the forward direction, we have the relation between the input field operator $\hat{s}_{\rm in}^{\dagger}(\omega)$, the output field operators $\hat{s}_{\rm out}^\dag(\omega)$ and $\hat{d}_{\rm out}^\dag(\omega)$, and the bath operators $\hat{A}_{{\rm out},i}^\dag(\omega)$,
\be
\hat{s}_{\rm in}^\dag(\omega)={\cal R}(\omega)\hat{s}_{\rm out}^\dag(\omega)+{\cal T}(\omega)\hat{d}_{\rm out}^\dag(\omega)+\sum_{i=1}^{n}{\cal S}_{i}(\omega)\hat{A}_{{\rm out},i}^\dag(\omega), \label{outdamp}
\ee 
where the index $s$ is dropped in the coefficients for ease of notation. The $i$-th nanoparticle loss coefficients, ${\cal S}_{i}(\omega)$, are also functions of the system parameters $g_{\rm in}$, $g_{\rm out}$, $g_{i,j}$ and $\omega_i$, and $|{\cal R}(\omega)|^2+|{\cal T}(\omega)|^2+\sum_{i=1}^{n} |{\cal S}_{i}(\omega)|^2=1$. This method again allows us to describe the nanoparticle array as an effective waveguide, with ${\cal T}(\omega) = |\tilde{\cal T}(\omega)|e^{i(kx\pm\pi)}$, where $|\tilde{\cal T}(\omega)|$ is the transmission in the ideal case (no damping) and the wavenumber $k=k^r+ik^i$ has become complex as a result of the damping~\cite{Loudon}, which now depends on the system parameters $g_{\rm in}$, $g_{\rm out}$, $g_{i,j}$, $\omega_i$ and the relaxation rates $\Gamma_i$ at each nanoparticle. Thus, we have that $|{\cal T}(\omega)|=|\tilde{\cal T}(\omega)|e^{-k^ix}$.

In Fig.~\ref{figresonancedamp}~(a)-(e) we plot the amplitude squared of the transmission, $|{\cal T}(\omega)|^2$, as the frequency is varied for an array of $n=1,~2,~3,~5$ and 7 nanoparticles respectively. To compare the damping with the ideal case shown in Fig.~\ref{figresonance}, we use the same system parameters: all local frequencies are equal $\omega_i=\omega_0,\forall i$, the couplings are equal $g_{i,j}=g_{\rm np}=-0.1 \omega_0, \forall i$ and its nearest neighbors $j$, and the source and drain couplings are set as $g_{\rm in/out}=0.01\omega_0$. In Appendix F we provide the analytical form for the ${\cal T}(\omega)$'s with damping. The damping rate for each nanoparticle depends on its size and is given by Matthiessen's rule~\cite{Brong}: $\Gamma=v_F/\lambda_B+v_F/\tilde{R}$, where for silver $\lambda_B=57$nm is the bulk mean-free path of an electron, $v_F=1.38 \times 10^6{\rm m}/{\rm s}$ is the velocity at the Fermi surface, and the effective radius $\tilde{R} \sim R$. We use $\Gamma_i=0.0158 \omega_0, \forall i$, which corresponds approximately to the damping rate for a silver nanoparticle with a radius $R$ in the range $20-100$nm. For a given $n$, the transmission spectral profiles in Fig.~\ref{figresonancedamp}~(a)-(e) again have $n$ resonances at frequencies $\omega_{r_j}=\omega_0+2g \cos (k^{r,j} d)$, where $k^{r,j}=j\pi/(n+1)d$ for $j=1,\dots,n$. However, the width of the resonances has been broadened and the height lowered as a result of the damping. In Fig.~\ref{figresonancedamp}~(f)-(j) we plot the real part of the effective wavenumber $k^r$ from Eq.~(\ref{knp}) for $n=1,~2,~3,~5$ and 7 respectively. Note that Eq.~(\ref{knp}) remains valid, as the imaginary part of the wavenumber $k$ is absorbed into the magnitude of the transmission, $|{\cal T}(\omega)|$. Points corresponding to the $k^{r,j}$ transmission resonance peaks from Fig.~\ref{figresonancedamp}~(a)-(e) are marked. Also included in these figures, as before, is the dispersion relation for the infinite array case (dashed line). 
\begin{figure*}[t]
\centering
\includegraphics[width=14cm]{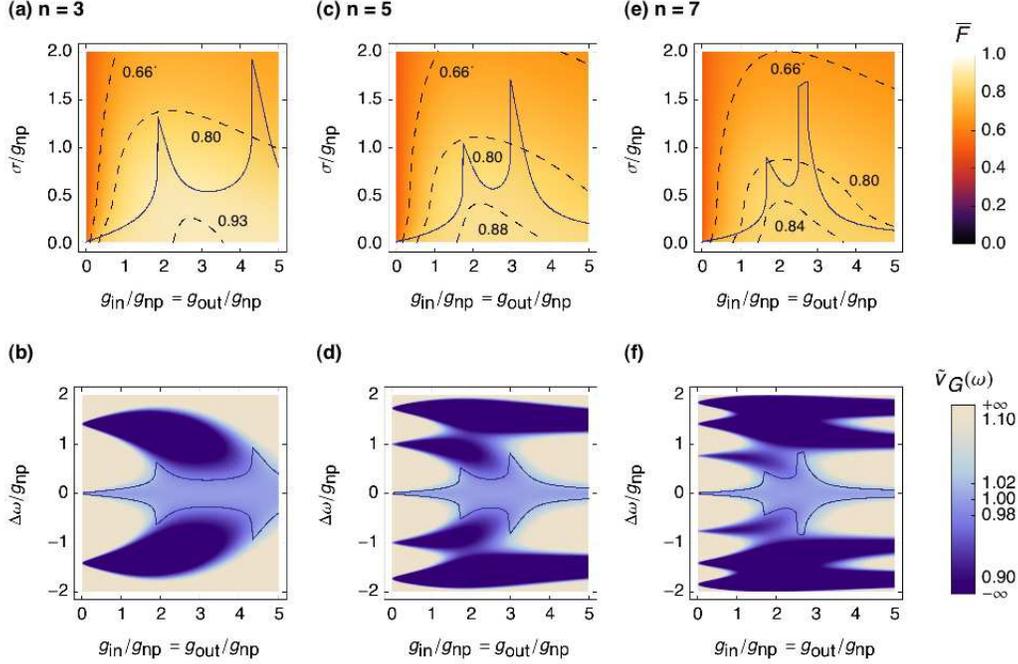}
\caption{Fidelity of quantum state transfer for a single qubit wavepacket under damping and scaled group velocity over a range of frequencies and couplings. {\bf (a):} Average fidelity $\bar{F}$ for transferring a qubit wavepacket over $n=3$ nanoparticles as the bandwidth ($\sigma$) and in/out couplings $g_{\rm in}=g_{\rm out}$ are modified. All parameters are scaled by $g_{\rm np}$ {\bf (b):} Scaled group velocity $\tilde{v}_G(\omega)=v_G(\omega)/v_G(\omega_0)$ as it deviates from that at the resonance frequency $\omega_0$ for $n=3$, showing the regions of approximate linear dispersion. {\bf (c):} Average fidelity for transferring a qubit wavepacket over $n=5$ nanoparticles. {\bf (d):} Scaled group velocity $\tilde{v}_G(\omega)$ as it deviates from that at the resonance frequency $\omega_0$ for $n=5$. {\bf (e):} Average fidelity for transferring a qubit wavepacket over $n=7$ nanoparticles. {\bf (f):} Scaled group velocity $\tilde{v}_G(\omega)$ as it deviates from that at the resonance frequency $\omega_0$ for $n=7$. In all group velocity plots, the region inside the blue lines for $\Delta \omega$ corresponds to the frequency range below the blue lines for $\sigma$ shown in the average fidelity plots.}
\label{fidloss}
\end{figure*}

From Fig.~\ref{figresonancedamp}~(a)-(e) one can clearly see that the transmission peaks are much reduced from the ideal values. However, despite this, it is possible to increase the maximum peak value by increasing the source and drain couplings, as shown in Fig.~\ref{arbdamp}~(a),~(b) and (c), where we plot a cross-section of the amplitude squared of the transmission, $|{\cal T}(\omega)|^2$, as the frequency $\omega$ and coupling $g_{\rm in}~(=g_{\rm out})$ are varied for an array of $n=3,~5$ and 7 nanoparticles respectively. Here, $\Delta \omega = \omega-\omega_0$ and as before, all parameters are in units of the nanoparticle coupling $g_{i,j}=g_{\rm np},~\forall i$ and its nearest neighbors $j$. One can see from Fig.~\ref{arbdamp} that as the source and drain couplings ($g_{\rm in}$ and $g_{\rm out}$) increase, the transmission maximum can be increased, although ultimately the damping dominates the transmission as $n$ increases, as can be seen by comparing Fig.~\ref{arbdamp}~(a) with (c).

\subsection{Qubit transfer}

We now discuss the fidelity of state transfer for a single qubit wavepacket state under realistic conditions of loss at each of the nanoparticles. After including the bath modes at each of the nanoparticles, one finds that the expression for the fidelity given in Eq.~(\ref{fidout}) remains valid (see Appendix C), with the fidelity depending only on the absolute value of transmission coefficient. In Fig.~\ref{fidloss}~(a),~(c) and (e) we show the average fidelity $\bar{F}$ for an array of $n=3$, 5 and 7 nanoparticles. The dashed lines correspond to fidelity contours with the lowest curve ($0.66^\cdot$) corresponding to the classical threshold for a quantum channel, as before. The solid blue curves bound a region (from below) in which the dispersion is approximately linear, $v_G(\omega)\simeq v_G(\omega_0)$. In Fig.~\ref{fidloss}~(b),~(d) and (f) we show the corresponding scaled group velocity $\tilde{v}_G(\omega)=v_G(\omega)/v_G(\omega_0)$ for $n=3$, 5 and 7 nanoparticles. For $n=3$, one can see in Fig.~\ref{fidloss}~(a) that the nanoparticle array can provide a transfer channel giving an average fidelity of up to $\sim0.93$ even when damping is present, in which for large bandwidths $\sigma$ the source and drain couplings $g_{\rm in}$ and $g_{\rm out}$ must be increased to values close to the limit of the weak coupling approximation, $|g_{\rm in,out}/g^{\rm max}_{\rm np}|=1$. Note that in these plots one cannot decrease the coupling $g_{\rm np}$ in order to reach $g_{\rm in,out}$ values much larger than 1, as we have set $g_{\rm np}=-0.1\omega_0$, unlike the ideal case where it could be modified. The reason for this restriction is that reducing the nanoparticle coupling $g_{\rm np}$ means the damping rates begin to dominate, lowering the maximum transmission and average fidelities further as a result. 

For $n=5$, one can see in Fig.~\ref{fidloss}~(c) that the maximum average fidelity attainable is $\sim0.88$; no contour can be plotted for 0.9 or above, regardless of the bandwidth $\sigma$. For $n=7$ and above, this situation then becomes gradually worse and one can see in Fig.~\ref{fidloss}~(e) that although the maximum average fidelity attainable is $\sim0.84$, the source and drain couplings need to be increased close to the weak coupling limit, in addition to the use of a narrow enough bandwidth. 
\begin{figure*}[t]
\centering
\includegraphics[width=17cm]{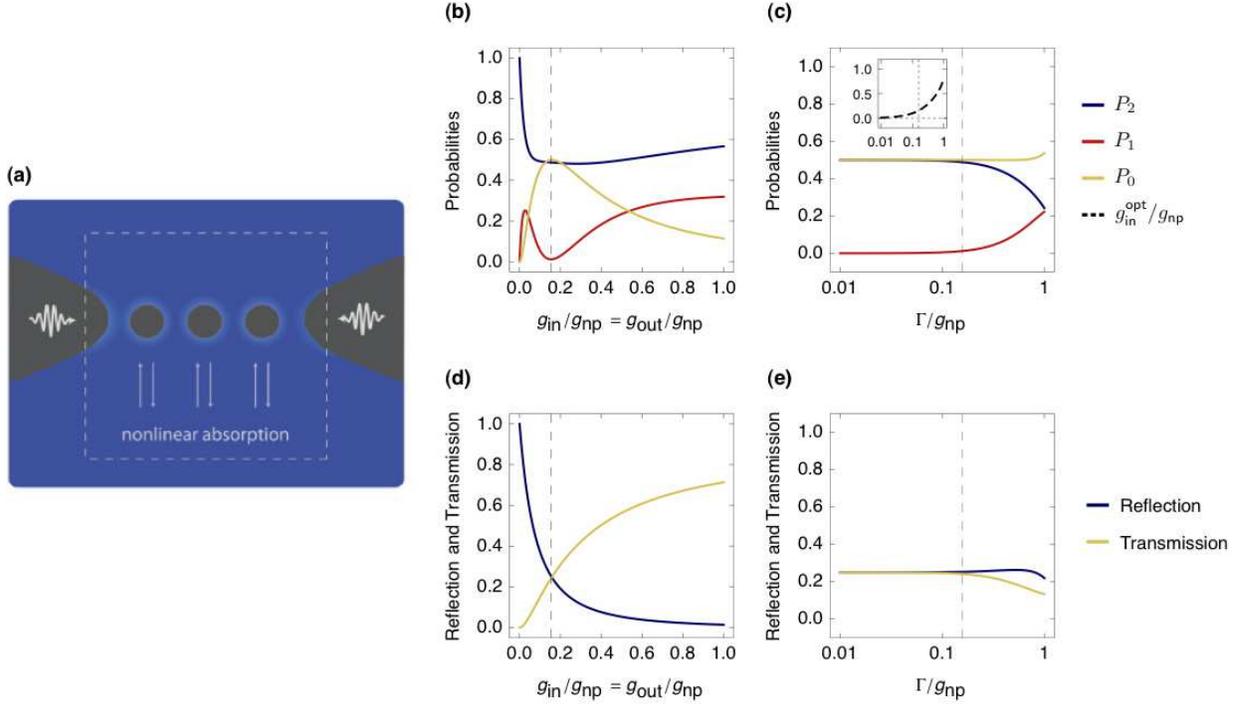}
\caption{(a): Plasmon interference. Here plasmons enter the nanoparticle array from both sides. One plasmon from the source and another from the drain. Nonlinear absorption occurs via quantum interference, even though the damping in the array is linear.
(b): Survival probabilities of zero-plasmon (yellow), one-plasmon (red), and two-plasmon (blue) for $n=3$, as the coupling $g_{\rm in}~(=g_{\rm out})$ is varied for $\Delta \omega=0$. (c): Survival probabilities for $n=3$ when the one-plasmon survival probability is minimized as the loss $\Gamma$ is varied for $\Delta \omega=0$. The values of $g_{\rm in}~(=g_{\rm out})$ at which minimization occurs are shown in the inset. (d) and (e): Reflection and transmission coefficients corresponding to the couplings in (b) and (c) respectively.}
\label{twophoton}
\end{figure*}

The results obtained here indicate that only small-sized arrays with $n \lesssim 7$ are useful for the transmission of qubit states encoded into the number state degree of freedom. However, it may be the case that for particular applications, short-distance communication ($\lesssim \mu$m) is required at optical frequencies, making the use of a nanoparticle array quite beneficial. For example, the nanoparticle waveguide could be used as an enhanced mediator between emitter systems on a very small scale. On the other hand, additional degrees of freedom for the LSP excitations, the embedding of emitter systems into the waveguides, novel types of metals with reduced damping rates and new schemes for achieving gain in plasmonic media may enable one to eventually counter the effects of loss highlighted here.

\subsection{Plasmon interference}

In Section III B we showed that in order to correctly describe the transfer of a quantum state through a metal nanoparticle array one requires the quantum formalism we have developed in this paper. Here, as an additional example of the necessity of a quantum formalism for the metal nanoparticle array, we investigate the interference of two plasmons. We consider the plasmons enter the array from opposite ends, one from the source and the other from the drain nanowire. The input state at the nanotips in this case can be written as
\be
\ket{\psi}_{\rm in}=\int_{-\infty}^{\infty} {\rm d}\omega_{s}\int_{-\infty}^{\infty} {\rm d}\omega_{d} \psi(\omega_{s},\omega_{d}) \hat{s}_{\rm in}^{\dagger}(\omega_{s})  \hat{d}_{\rm in}^{\dagger}(\omega_{d}) \ket{0}_{s,d,A},\nonumber
\ee
where $\ket{0}_{s,d,A}$ denotes the vacuum state for the source, drain and baths and the normalization of the state vector imposes a nomalization on $\psi(\omega_{s},\omega_{d})$, so that $\int_{-\infty}^{\infty} {\rm d}\omega_{s}\int_{-\infty}^{\infty} {\rm d}\omega_{d} | \psi(\omega_{s},\omega_{d})|^{2}=1$.
By using the scattering matrix given in Eq.~(\ref{outdamp2}) of Appendix E, that describes forward and backward propagation in the array, we have the output
\bqa\label{outputstate}
\ket{\psi}_{\rm out}&=& \nonumber \\
&& \hskip-1cm \int_{-\infty}^{\infty} {\rm d}\omega_{s}\int_{-\infty}^{\infty} {\rm d}\omega_{d} \psi(\omega_{s},\omega_{d}) \times\nonumber\\
&&
\hskip-1cm \bigg({{\cal R}_{s}}(\omega_{s})\hat{s}_{\rm out}^\dag(\omega_{s})+{{\cal T}_{s}}(\omega_{s})\hat{d}_{\rm out}^\dag(\omega_{s})+ \hat{F}^{\dagger}_{s}(\omega_{s}) \bigg) \times \nonumber \\
&&
\hskip-1cm \bigg({{\cal T}_{d}}(\omega_{d})\hat{s}_{\rm out}^\dag(\omega_{d})+{{\cal R}_{d}}(\omega_{d})\hat{d}_{\rm out}^\dag(\omega_{d})+\hat{F}^{\dagger}_{d}(\omega_{d}) \bigg)
\ket{0}_{s,d,A}, \nonumber 
\eqa
where the noise operators are defined as $\hat{F}^{\dagger}_{s}(\omega_{s})=\sum_i{{\cal S}}_{s,i}(\omega_{s})\hat{A}_{{\rm out},i}^\dag(\omega_{s})$ and $\hat{F}^{\dagger}_{d}(\omega_{d})=\sum_i{{\cal S}}_{d,i}(\omega_{d})\hat{A}_{{\rm out},i}^\dag(\omega_{d})$. 
We consider small bandwidths for $\psi(\omega_{s},\omega_{d})$ over which the transmission, reflection and damping coefficients do not vary appreciably and therefore these coefficients will be approximated as frequency independent. The case of $g_{\rm in}=g_{\rm out}$ is considered, so that we have ${\cal T}_{d}(\omega)={\cal T}_{s}(\omega)={\cal T}$ and ${\cal R}_d(\omega)={\cal R}_s(\omega)={\cal R}$. The probability of finding two plasmons in the source nanowire and none in the drain nanowire is then (see Appendix G)
\bqa
P(2_{s}, 0_{d}) = |{\cal R} |^{2} |{\cal T}|^{2} (1+ {\cal I}), \label{p20}
\eqa
where we have introduced the (real) overlap integral
\be
{\cal I}=\int_{-\infty}^{\infty} {\rm d}\omega_s \int_{-\infty}^{\infty} {\rm d}\omega_d \psi(\omega_s,\omega_d) \psi^{*}(\omega_d,\omega_s).
\ee
Here, unit quantum efficiency of the photon detector and infinite counting time are assumed. Similarly, the remaining nonzero probabilities are
\bqa \label{probphotons}
P(0_{s}, 2_{d}) &=& P(2_{s}, 0_{d}), \nonumber\\
P(1_{s}, 1_{d}) &=& |{\cal R}|^{4} + |{\cal T}|^{4} +({\cal R}^{2}{\cal T}^{*2}+{\cal R}^{*2}{\cal T}^{2}){\cal I}, \nonumber\\
P(1_{s}, 0_{d}) &=& (|{\cal R}|^{2}+|{\cal T}|^{2})(1-|{\cal R}|^{2}-|{\cal T}|^{2}) \nonumber \\
&& \hskip0.5cm -({\cal R}{\cal T}^{*}+{\cal R}^{*}{\cal T})^{2}{\cal I}, \nonumber\\
P(0_{s}, 1_{d})&=&P(1_{s}, 0_{d}), \nonumber\\
P(0_{s}, 0_{d}) &=& (1-|{\cal R}|^{2}-|{\cal T}|^{2})^{2} + ({\cal R}^{*}{\cal T}+{\cal T}^{*}{\cal R})^{2}{\cal I}. \nonumber\\
\eqa
Here, for simplicity, we consider that the plasmons have the same wavepacket profile, \textit{i.e.}, $\psi(\omega_{s},\omega_{d})=\xi(\omega_{s})\xi(\omega_{d})$, where $\xi(\omega)$ is given in Eq.~(\ref{gaussian}). If ${\cal I}=0$, the Fourier transform of the spectral amplitudes for the two plasmons do not overlap in time in the nanoparticle array, and the probabilities in Eqs.~(\ref{p20})~and~(\ref{probphotons}) describe the case of two independent particles~\cite{Barnett}. On the other hand, if ${\cal I}=1$, the Fourier transform of the amplitudes overlap perfectly in time. In this case, temporal and spectral indistinguishabilities are immediately satisfied and for $|{\cal R} |^{2}= |{\cal T}|^{2}=1/2$ one recovers the well-known Hong-Ou-Mandel (HOM) quantum interference effect~\cite{HOM}, where the two excitations are always found to be in the same output mode: $P(2_{s}, 0_{d})=P(0_{s}, 2_{d})=1/2$, with all others being zero. In general, however, when $|{\cal R} |^{2}\neq |{\cal T}|^{2}$ or damping is present ($|{\cal R} |^{2}+|{\cal T}|^{2} < 1$), the probabilities for two, one, or no plasmons to survive are $P_{2}=P(2_{s}, 0_{d})+P(0_{s}, 2_{d})+P(1_{s}, 1_{d})$, $P_{1}=P(1_{s}, 0_{d})+P(0_{s}, 1_{d})$, and $P_{0}=P(0_{s}, 0_{d})$, respectively, with $\sum_{i=0}^{2}P_i=1$. 

For an array of $n=3$ nanoparticles, we plot in Fig.~\ref{twophoton}~(b) the survival probabilities as the coupling $g_{\rm in}~(=g_{\rm out})$ is varied for $\Delta \omega=0$, $\Gamma=0.0158 \omega_{0}$, and $g_{\rm np}=-0.1\omega_{0}$. One can see that the probability for one of the plasmons to survive (or be absorbed), $P_1$, can be very low depending on the in/out coupling. Indeed, at a particular point marked by the dashed line, \textit{nonlinear absorption} occurs: either both plasmons are absorbed, $P_0\sim1/2$, or neither is absorbed, $P_2\sim1/2$, even though the damping in the nanoparticle array is a linear process. Surprisingly there is no one-plasmon absorption, $P_1\sim 0$. This effect is due to quantum interference of the plasmons and cannot be described in terms of a classical treatment of the nanoparticle array~\cite{Barnett}. In Fig.~\ref{twophoton}~(d) we show the corresponding reflection, $|{\cal R} |^{2}$, and transmission, $|{\cal T} |^{2}$, coefficients. The transmission coefficient in this plot can also be seen by taking a cross-section from Fig.~\ref{arbdamp}~(a) at $\Delta \omega=0$. One can see in Fig.~\ref{twophoton}~(d) that the nonlinear absorption effect is maximized at a similar point to that for the HOM interference effect: reflection and transmission coefficients are equalized, but at $1/4$ instead of $1/2$ due to the necessary presence of damping in order to see nonlinear absorption~\cite{Barnett}. 

In Fig.~\ref{twophoton}~(c), we show how increasing the loss at each nanoparticle affects the two-plasmon interference for $n=3$. Here, $P_{1}$ is minimized by modifying $g_{\rm in}$ ($=g_{\rm out}$) as the loss $\Gamma$ is increased for $\Delta \omega=0$. The corresponding $P_0$ and $P_2$ are also shown. One can see that nonlinear absorption can be made to occur over a large range of loss. The values of $g_{\rm in}~(=g_{\rm out})$ at which $P_1$ is minimized are shown in the inset and the corresponding reflection and transmission coefficients are shown in Fig.~\ref{twophoton}~(e). Note that as the amount of loss increases, both the minimum value of $P_{1}$ and the required coupling $g_{\rm in}~(=g_{\rm out})$ are increased also. In particular, one can see in Fig.~\ref{twophoton}~(e), that as the damping in the array increases, it becomes impossible to equalize the reflection and transmission coefficients by changing $g_{\rm in}$, as the transmission is affected more by loss within the array. This asymmetry leads to an eventual breakdown of the quantum interference effect and subsequently the nonlinear absorption.

The behaviour shown in Fig.~\ref{twophoton}~(c) allows us to predict the growing trend of the minimum value of $P_{1}$ and the optimal value of $g_{\rm in}~(=g_{\rm out})$ as $n$ increases. This is because the overall amount of loss in the array effectively increases as the number of nanoparticles is increased. For $\Delta \omega=0$, $\Gamma=0.0158 \omega_{0}$ and $g_{\rm np}=-0.1\omega_{0}$, one finds that $P^{\rm min}_{1}=0.012,~0.034$ and $0.063$ when $g^{\rm opt}_{\rm in}/g_{\rm np}=0.1543,~0.2223$ and $0.2824$ for $n=3,~5$ and $7$, respectively. The corresponding zero and two-plasmon probabilities are $P_0=0.4999,~0.4995$ and $0.4990$, and $P_2=0.4880,~0.4663$ and $0.4380$. Thus, nonlinear absorption by two-plasmon interference is present in the nanoparticle array for $n=3, 5$ and $7$. The nanoparticle array may therefore act as an effective two-plasmon absorber, despite the linear optical properties assumed in the model. 

\section{Summary}

In this work we studied the use of an array of metallic nanoparticles as a channel for on-chip nanophotonic quantum communication. After introducing the model for the physical system in the quantum regime, the transfer of a quantum state encoded in the form of a single-qubit wavepacket was studied under ideal conditions. We then showed the necessity for our quantum formalism in predicting the outcomes of measurable physical observables. The effects of loss in the metal were included in our study, thus putting the investigation into a more practical setting and allowing the quantification of the performance of realistic nanoparticle arrays as quantum channels. For this task we used the average fidelity for the state transfer. We found that small-sized arrays are practically useful for the transmission of qubit states encoded into the number state degree of freedom. We also showed that nonlinear absorption can occur by quantum interference, where two plasmons are absorbed or neither is absorbed. Thus, the nanoparticle array can act as an effective two-plasmon absorber, and the observation of this quantum interference effect may open up new kinds of plasmonic interference experiments in the quantum domain. Our study highlights the benefits as well as the drawbacks associated with nanophotonic periodic quantum systems that use surface plasmons. The techniques introduced in this work may assist in the further theoretical and experimental study of plasmonic nanostructures for quantum control applications and probing nanoscale optical phenomena.

\acknowledgments 
We thank Prof.~M.~S.~Kim, Dr.~S.~K.~Ozdemir and Prof.~J.~Takahara for discussions. This work was supported by the UK's Engineering and Physical Sciences Research Council (EPSRC) and the National Research Foundation (NRF) of Korea grant funded by the Korea Government (Ministry of Education, Science and Technology; grant numbers 2010-0015059 and 3348-20100018).

\renewcommand{\theequation}{A-\arabic{equation}}
\setcounter{equation}{0}  
\section*{APPENDIX A}  

Here we use input-output formalism~\cite{CG,WM} for the nanoparticle array to obtain an effective scattering matrix. We start with the Heisenberg equation of motion for an operator $\hat{O}$, given by $\frac{d\hat{O}}{dt}=-\frac{i}{\hbar}[\hat{O},\hat{H}]$, and substitute the Hamiltonian $\hat{H}$ in Eq.~(\ref{HamT}) to obtain the equations of motion for each of the system operators
\bqa
\frac{d\hat{s}(\omega)}{dt}&=&-i\omega \hat{s}(\omega) + g_{\rm in}(\omega)\hat{a}_1,  \label{eomin} \\
\frac{d\hat{a}_1}{dt}&=&-\frac{i}{\hbar}[\hat{a}_1,\hat{H}_{np}]-\int_{-\infty}^{\infty}d\omega g_{\rm in}(\omega)\hat{s}(\omega),  \label{eom1} \\
\frac{d\hat{a}_i}{dt}&=&-\frac{i}{\hbar}[\hat{a}_i,\hat{H}_{np}],~~~i=2,\dots, n-1,   \label{eomi} \\
\frac{d\hat{a}_n}{dt}&=&-\frac{i}{\hbar}[\hat{a}_n,\hat{H}_{np}]-\int_{-\infty}^{\infty}d\omega g_{\rm out}(\omega)\hat{d}(\omega),  \label{eomn} \\
\frac{d\hat{d}(\omega)}{dt}&=&-i\omega \hat{d}(\omega) + g_{\rm out}(\omega)\hat{a}_n. \label{eomout}
\eqa
Here we have introduced an explicit time dependence in the frequency space operators $\hat{s}(\omega)$ and $\hat{d}(\omega)$ in the source and drain respectively. This is because the internal field of the nanoparticle array may acquire some non-trivial dynamics which forces the external fields in the source and drain to have a time dependence that is different from the free field dynamics~\cite{CG,WM}. With the above set of coupled equations of motion we find boundary conditions for the system before proceeding to solve them. Using the following solutions for the first and last equations (Eqs.~(\ref{eomin}) and (\ref{eomout}))
\bqa
\hat{s}(\omega)&=&e^{-i \omega(t-t_0)}\hat{s}_0(\omega)+g_{\rm in}(\omega)\int_{t_0}^{t}e^{-i \omega(t-t')}\hat{a}_1(t')dt' \nonumber, \\
\hat{d}(\omega)&=&e^{-i \omega(t-t_0)}\hat{d}_0(\omega)+g_{\rm out}(\omega)\int_{t_0}^{t}e^{-i \omega(t-t')}\hat{a}_n(t')dt', \nonumber
\eqa
where $t_0 < t$, with $\hat{s}_0(\omega)$ and $\hat{d}_0(\omega)$ as the operators for $\hat{s}(\omega)$ and $\hat{d}(\omega)$ respectively at time $t=t_0$ as {\it initial} boundary conditions, one finds the equations of motion (Eqs.~(\ref{eom1}) and (\ref{eomn})) for the first and last nanoparticle become
\bqa
\frac{d\hat{a}_1}{dt}&=&-\frac{i}{\hbar}[\hat{a}_1,\hat{H}_{np}]-\frac{g_{\rm in}}{2}\hat{a}_1+\sqrt{g_{\rm in}} \hat{s}_{\rm in}, \label{sini} \\
\frac{d\hat{a}_n}{dt}&=&-\frac{i}{\hbar}[\hat{a}_n,\hat{H}_{np}]-\frac{g_{\rm out}}{2}\hat{a}_n+\sqrt{g_{\rm out}} \hat{d}_{\rm in}, \label{dini}   
\eqa
where we have defined the input field operators as $\hat{s}_{\rm in}(t)=-(2 \pi)^{-1/2} \int_{-\infty}^{\infty}d\omega e^{-i \omega(t-t_0)}\hat{s}_0(\omega)$
and $\hat{d}_{\rm in}(t)=-(2 \pi)^{-1/2} \int_{-\infty}^{\infty}d\omega e^{-i \omega(t-t_0)}\hat{d}_0(\omega)$. Here we have assumed the couplings $g_{\rm in}(\omega)$ and $g_{\rm out}(\omega)$ are constant over a band of frequencies about the characteristic excitation frequency being considered, $g_{\rm in}^2(\omega)=g_{\rm in}/2 \pi$ and $g_{\rm out}^2(\omega)=g_{\rm out}/2 \pi$. This assumption is valid for negligible change in the similarity of the modefunction profiles at the tip and nanoparticles over the bandwidth. We assume this can be achieved given a narrow enough band of frequencies along with an optimized nanotip geometry. 

Using alternative solutions for the first and last equations of motion
\bqa
\hat{s}(\omega)&=&e^{-i \omega(t-t_1)}\hat{s}_1(\omega)-g_{\rm in}(\omega)\int_{t}^{t_1}e^{-i \omega(t-t')}\hat{a}_1(t')dt' \nonumber, \\
\hat{d}(\omega)&=&e^{-i \omega(t-t_1)}\hat{d}_1(\omega)-g_{\rm out}(\omega)\int_{t}^{t_1}e^{-i \omega(t-t')}\hat{a}_n(t')dt', \nonumber
\eqa
where $t_1 > t$, with $\hat{s}_1(\omega)$ and $\hat{d}_1(\omega)$ as the operators for $\hat{s}(\omega)$ and $\hat{d}(\omega)$ respectively at time $t=t_1$ as {\it final} boundary conditions, one finds the equations of motion (Eqs.~(\ref{eom1}) and (\ref{eomn})) for the first and last nanoparticle become
\bqa
\frac{d\hat{a}_1}{dt}&=&-\frac{i}{\hbar}[\hat{a}_1,\hat{H}_{np}]+\frac{g_{\rm in}}{2}\hat{a}_1-\sqrt{g_{\rm in}} \hat{s}_{\rm out},  \label{sfin} \\
\frac{d\hat{a}_n}{dt}&=&-\frac{i}{\hbar}[\hat{a}_n,\hat{H}_{np}]+\frac{g_{\rm out}}{2}\hat{a}_n-\sqrt{g_{\rm out}} \hat{d}_{\rm out},  \label{dfin} 
\eqa
where we have defined the output field operators as $\hat{s}_{\rm out}(t)=(2 \pi)^{-1/2} \int_{-\infty}^{\infty}d\omega e^{-i \omega(t-t_1)}\hat{s}_1(\omega)$
and $\hat{d}_{\rm out}(t)=(2 \pi)^{-1/2} \int_{-\infty}^{\infty}d\omega e^{-i \omega(t-t_1)}\hat{d}_1(\omega)$.

Taking Eq.~(\ref{sfin}) and subtracting Eq.~(\ref{sini}) gives the boundary condition
\be
\hat{a}_1(t)=\frac{1}{\sqrt{g_{\rm in}}}(\hat{s}_{\rm in}(t)+\hat{s}_{\rm out}(t)). \label{bound1}
\ee
Similarly, taking Eq.~(\ref{dfin}) and subtracting Eq.~(\ref{dini}) gives the boundary condition
\be
\hat{a}_n(t)=\frac{1}{\sqrt{g_{\rm out}}}(\hat{d}_{\rm in}(t)+\hat{d}_{\rm out}(t)). \label{bound2}
\ee
Note that throughout we assume the dispersion is negligible for the initial/final excitations of the source and drain fields. In this sense we are interested only in the relation between the input/output propagating fields in the nanowires near the tips. The dispersion during propagation of the excitations in the nanowires can be incorporated into the model by using standard methods~\cite{Loudon}. On the other hand, the dispersion in the array is included in the model automatically, although we will need to ensure later that minimal broadening of the bandwidth due to dispersion occurs during the propagation for the relation $g_{\rm out}^2(\omega)=g_{\rm out}/2\pi$ to still hold. We will see that this is a reasonable assumption for small-sized arrays.

We use the relations $[\hat{a}_1,\hat{H}_{np}]=\hbar \omega_1 \hat{a}_1+\hbar g_{1,2} \hat{a}_2$, $[\hat{a}_n,\hat{H}_{np}]=\hbar \omega_n \hat{a}_n+\hbar g_{n-1,n} \hat{a}_{n-1}$ and $[\hat{a}_i,\hat{H}_{np}]=\hbar \omega_i \hat{a}_i+\hbar g_{i-1,i} \hat{a}_{i-1}+\hbar g_{i,i+1} \hat{a}_{i+1}$ (for $i=2,\dots, n-1$) as well as defining the Fourier components of the nanoparticle field operators as $\hat{a}_i(t)=(2 \pi)^{-1/2}\int_{-\infty}^{\infty}e^{-i\omega t}\hat{a}_i(\omega)d\omega,~\forall i$, and we rewrite the source/drain operators as $\hat{s}_{\rm in}(t)=(2 \pi)^{-1/2}\int_{-\infty}^{\infty}e^{-i\omega t}\hat{s}_{\rm in}(\omega)d\omega$ and $\hat{d}_{\rm in}(t)=(2 \pi)^{-1/2}\int_{-\infty}^{\infty}e^{-i\omega t}\hat{d}_{\rm in}(\omega)d\omega$ ($\hat{s}_{\rm in}(\omega)$ and $\hat{d}_{\rm in}(\omega)$ are general spectral operators). Then, we find upon substitution into Eqs.~(\ref{eomi}), (\ref{sini}) and (\ref{dini}) the following set of coupled equations for the frequency operators
\bqa
\left[i(\omega-\omega_1)-\frac{g_{\rm in}}{2}\right] \hat{a}_1(\omega)&=& i g_{1,2}\hat{a}_2(\omega)-\sqrt{g_{\rm in}}\hat{s}_{\rm in}(\omega), \nonumber \\
i(\omega-\omega_i) \hat{a}_i(\omega)&=& i g_{i-1,i}\hat{a}_{i-1}(\omega) + i g_{i+1,i}\hat{a}_{i+1}(\omega), \nonumber \\
&&\qquad \qquad{\rm for}~i=2,\dots, n-1, \nonumber \\
\left[i(\omega-\omega_n)-\frac{g_{\rm out}}{2}\right] \hat{a}_n(\omega)&=& i g_{n-1,n}\hat{a}_{n-1}(\omega)-\sqrt{g_{\rm out}}\hat{d}_{\rm in}(\omega) , \nonumber
\eqa 
as well as boundary conditions from Eqs.~(\ref{bound1}) and (\ref{bound2})
\bqa
\hat{a}_1(\omega)&=&\frac{1}{\sqrt{g_{\rm in}}}(\hat{s}_{\rm in}(\omega)+\hat{s}_{\rm out}(\omega)), \nonumber \\
\hat{a}_n(\omega)&=&\frac{1}{\sqrt{g_{\rm out}}}(\hat{d}_{\rm in}(\omega)+\hat{d}_{\rm out}(\omega)). \nonumber
\eqa
Using the above set of coupled equations we can eliminate the internal nanoparticle operators $\hat{a}_i$~\cite{CG,WM} to obtain
\bqa
\hat{s}_{\rm in}(\omega)&=& {\cal R}_s^*(\omega)\hat{s}_{\rm out}(\omega)+{\cal T}^*_s(\omega)\hat{d}_{\rm out}(\omega), \nonumber\\
\hat{d}_{\rm in}(\omega)&=& {\cal T}^*_d(\omega)\hat{s}_{\rm out}(\omega)+{\cal R}^*_d(\omega)\hat{d}_{\rm out}(\omega), \nonumber
\eqa 
where the transmission ${\cal T}_{s,d}$ and reflection ${\cal R}_{s,d}$ coefficients are functions of the system parameters $g_{\rm in}$, $g_{\rm out}$, $g_{i,j}$ and $\omega_i$, and the relation $|{\cal R}_{s,d}(\omega)|^2+|{\cal T}_{s,d}(\omega)|^2=1$ holds.

\renewcommand{\theequation}{B-\arabic{equation}}
\setcounter{equation}{0}  
\section*{APPENDIX B}  
Here we provide the analytical forms for the ${\cal T}(\omega)$'s for $n=1,~2,~3,~5$ and 7 nanoparticles respectively (no damping), where we have set $g_{\rm out}=g_{\rm in}$,
\bqa
{\cal T}_1(\omega)&=&\frac{g_{\rm in}}{g_{\rm in}-i (\omega-\omega_0)}, \nonumber \\
{\cal T}_2(\omega)&=&\frac{-4ig_{\rm np} g_{\rm in}}{4g_{\rm np}^2+(g_{\rm in}-2i (\omega-\omega_0))^2}, \nonumber \\
{\cal T}_3(\omega)&=&-4g_{\rm np}^2 g_{\rm in}[(g_{\rm in}-2i (\omega-\omega_0))(4g_{\rm np}^2 \nonumber \\
&& \qquad \qquad \qquad -(\omega-\omega_0)(i g_{\rm in}+2(\omega-\omega_0)))]^{-1}, \nonumber \\
{\cal T}_5(\omega)&=&4g_{\rm np}^4 g_{\rm in}[(2g_{\rm np}^2(g_{\rm in}-3i(\omega-\omega_0))        \nonumber \\
&& \qquad -(g_{\rm in}-2i(\omega-\omega_0))(\omega-\omega_0)^2) \times  \nonumber \\
&&\qquad \qquad (2g_{\rm np}^2-(\omega-\omega_0)(i g_{\rm in}+2(\omega-\omega_0))) ]^{-1}, \nonumber \\
{\cal T}_7(\omega)&=&-4g_{\rm np}^6 g_{\rm in}[(g_{\rm np}^2(g_{\rm in}-4i(\omega-\omega_0)) \nonumber \\
&&\qquad -(g_{\rm in}-2i(\omega-\omega_0))(\omega-\omega_0)^2) \times  \nonumber \\
&&\qquad \qquad(4g_{\rm np}^4+(\omega-\omega_0)^3(i g_{\rm in}+2(\omega+\omega_0)) \nonumber \\
&&\qquad \qquad ~~-g_{\rm np}^2(\omega-\omega_0)(3ig_{\rm in}+8(\omega+\omega_0)))]^{-1}. \nonumber 
\eqa 

\renewcommand{\theequation}{C-\arabic{equation}}
\setcounter{equation}{0}  
\section*{APPENDIX C}  

Here we show how to obtain the output density matrix for the qubit state entering the drain nanowire. This is done in the general case of damping (see Section IV). To obtain the case of no loss, simply set ${\cal S}_i(\omega)=0,~\forall i$. Starting with the single qubit wavepacket in the input modes of the source nanowire
\be
\ket{\psi}_s=a\ket{0}_s+b \int_{-\infty}^{\infty} {\rm d}\omega \xi(\omega)\hat{s}^\dag_{\rm in}(\omega)\ket{0}_s.\nonumber
\ee
and making use of Eq.~(\ref{outdamp}) of Section IV, (equivalent to Eq.~(\ref{inop1}) of Section II, when ${\cal S}_i(\omega)=0,~\forall i$), substituting for $\hat{s}^\dag_{\rm in}(\omega)$ one obtains the state $\rho_{s,d,A}=\pro{\phi}{s,d,A}{\phi}$ which describes the total state in the external output modes of the {\it source-nanoparticle-drain} system, where
\bqa
\ket{\phi}_{s,d,A}&=&a\ket{0}_{s,d,A}+b \int_{-\infty}^{\infty} {\rm d}\omega \xi(\omega)\bigg[ {\cal R}(\omega)\hat{s}^\dag_{\rm out}(\omega)  \nonumber\\
&&\qquad +{\cal T}(\omega)\hat{d}^\dag_{\rm out}(\omega) +  \sum_{i=1}^n {\cal S}_i(\omega)\hat{A}^\dag_{\rm out,i}(\omega)\bigg]\ket{0}_{s,d,A} \nonumber
\eqa
Removing the source modes from the description of the state $\rho_{s,d,A}$ is achieved mathematically by {\it tracing} them out to give
\be
\rho_{d,A}=_s\hskip-0.1cm\bra{0}\rho_{s,d,A}\ket{0}_s+\int_{-\infty}^{\infty} {\rm d}\omega~ _s\hskip-0.1cm\bra{1_\omega}\rho_{s,d,A}\ket{1_\omega}_s.\nonumber
\ee
Tracing out the $i$ bath modes recursively in a similar way gives
\bqa
\rho_d&=&\bigg(|a|^2+|b|^2\int_{-\infty}^{\infty} {\rm d}\omega |\xi(\omega)|^2(1-|{\cal T}(\omega)|^2)\bigg) \pro{0}{d}{0} \nonumber \\
&& + a b^*\int_{-\infty}^{\infty} {\rm d}\omega \xi^*(\omega){\cal T}^*(\omega) \pro{0}{d}{1_{\omega}} \nonumber \\
&& +a^* b \int_{-\infty}^{\infty} {\rm d}\omega \xi(\omega){\cal T}(\omega) \pro{1_{\omega}}{d}{0} \nonumber \\
&&+|b|^2\int_{-\infty}^{\infty} {\rm d}\omega \int {\rm d}\omega'\xi(\omega)\xi^*(\omega') \times \nonumber \\
&& \qquad \qquad \qquad \qquad {\cal T}(\omega){\cal T}^*(\omega')\pro{1_{\omega}}{d}{1_{\omega'}}. \label{outstatApp}
\eqa

\renewcommand{\theequation}{D-\arabic{equation}}
\setcounter{equation}{0}  
\section*{APPENDIX D}  
Here we derive the fidelity for the coherent state transfer described in Section III B. First, we consider two continuous-mode coherent states defined as
\bqa
\ket{\{\alpha\}} &=& \mathrm{exp}(\hat{b}_{\alpha}^{\dagger}-\hat{b}_{\alpha}) \ket{0}, \nonumber\\
\ket{\{\beta\}} &=& \mathrm{exp}(\hat{b}_{\beta}^{\dagger}-\hat{b}_{\beta}) \ket{0},\nonumber
\eqa
where the photon wavepacket operators are given by $\hat{b}_{\alpha}^{\dagger} = \int_{-\infty}^{\infty} {\rm d}\omega \alpha(\omega)\hat{b}^\dag (\omega)$ and $\hat{b}_{\beta}^{\dagger} = \int_{-\infty}^{\infty} {\rm d}\omega \beta(\omega)\hat{b}^\dag (\omega)$, with $\int_{-\infty}^{\infty} {\rm d}\omega |\alpha(\omega)|^{2}=\bar{n}_{\alpha}$ and $\int_{-\infty}^{\infty} {\rm d}\omega |\beta(\omega)|^{2}=\bar{n}_{\beta}$. The operators $\hat{b}^{\dagger}(\omega)$ $(\hat{b}(\omega))$ represent the creation (annihilation) operators associated with a field excitation, which obey the bosonic commutation relation $[\hat{b}(\omega),\hat{b}^\dag(\omega')]=\delta (\omega-\omega')$. In general, the fidelity between the two continuous-mode coherent states, defined as $F=| \left<\{\beta\} |\{\alpha\} \right>|^{2}$, is found by direct substitution to be
\be
F= \mathrm{exp}[ - \int_{-\infty}^{\infty} {\rm d}\omega |\alpha(\omega)-\beta(\omega)|^{2} ].\nonumber
\ee
The fidelity between the transferred coherent state $\ket{\{ \alpha {\cal T} \}}$ in Section III B and the ideal transferred state $\ket{\{ \alpha' \}}$, where $\alpha'(\omega)=\alpha(\omega) e^{i k x}$, is then
\be
F= \mathrm{exp} \left[- \int_{-\infty}^{\infty} {\rm d}\omega |\alpha(\omega)|^{2}( |{\cal T}(\omega)|-1)^{2} \right].\nonumber
\ee

\renewcommand{\theequation}{E-\arabic{equation}}
\setcounter{equation}{0}  
\section*{APPENDIX E}  
Here we use input-output formalism for the nanoparticle array under realistic conditions of loss at each nanoparticle, where the damping is modeled by coupling of the field at each nanoparticle to an independent bath mode. The interaction of each nanoparticle to its bath mode takes the same form as the coupling of the first nanoparticle to the source nanowire tip, except with a coupling strength determined by the rate of damping to match the classical case, as done previously for the inter-particle couplings $g_{i,j}$. This approach assumes a weak damping rate and Markov approximation for the bath modes~\cite{CG,WM}. To mathematically incorporate the bath modes into our model, the original coupled equations are modified to become
\bqa
&& \left[i(\omega-\omega_1)-\frac{g_{\rm in}}{2}-\frac{\Gamma_1}{2}\right] \hat{a}_1(\omega)= i g_{1,2}\hat{a}_2(\omega)-\sqrt{g_{\rm in}}\hat{s}_{\rm in}(\omega)  \nonumber \\
&&\hskip4.6cm -\sqrt{\Gamma_1}\hat{A}_{{\rm in},1}(\omega),  \nonumber \\
&& \left[i(\omega-\omega_i)-\frac{\Gamma_i}{2}\right]\hat{a}_i(\omega)= i g_{i-1,i}\hat{a}_{i-1}(\omega) + i g_{i+1,i}\hat{a}_{i+1}(\omega) \nonumber \\
&&\hskip3.2cm -\sqrt{\Gamma_i}\hat{A}_{{\rm in},i}(\omega),~{\rm for}~i=2,\dots, n-1, \nonumber \\
&& \left[i(\omega-\omega_n)-\frac{g_{\rm out}}{2}-\frac{\Gamma_n}{2}\right] \hat{a}_n(\omega)= i g_{n-1,n}\hat{a}_{n-1}(\omega) \nonumber \\
&&\hskip3cm -\sqrt{g_{\rm out}}\hat{d}_{\rm in}(\omega)-\sqrt{\Gamma_n}\hat{A}_{{\rm in},n}(\omega), \nonumber 
\eqa
where $\Gamma_i$ corresponds to the electronic relaxation rate at nanoparticle $i$. Extra boundary conditions are then imposed on the system dynamics given by
\be
\hat{a}_i(\omega)=\frac{1}{\sqrt{\Gamma_i}}(\hat{A}_{{\rm in},i}(\omega)+\hat{A}_{{\rm out},i}(\omega)), ~{\rm for}~i=1,\dots, n
\ee
As in the case of the source and drain nanowire system, the bath operators $\hat{A}(\omega)$ obey the bosonic commutation relations $[\hat{A}(\omega),\hat{A}^\dag(\omega')]=\delta (\omega-\omega')$. By solving the above new set of coupled equations and eliminating the internal nanoparticle operators $\hat{a}_i$, we obtain a scattering matrix linking the source, drain and bath operators. Then for a given input state $\ket{\psi}$ from the source, we take the bath modes (and drain) to be initially in the vacuum state $\ket{0}$. The scattering matrix is applied and the bath modes are traced out to obtain the effective transmission in the array. In general one finds 
\bqa 
\hat{s}_{\rm in}^\dag(\omega)&=&{\cal R}_{s}(\omega)\hat{s}_{\rm out}^\dag(\omega)+{\cal T}_{s}(\omega)\hat{d}_{\rm out}^\dag(\omega)+\sum_i{\cal S}_{s,i}(\omega)\hat{A}_{{\rm out},i}^\dag(\omega) , \nonumber\\
\hat{d}_{\rm in}^\dag(\omega)&=&{\cal T}_{d}(\omega)\hat{s}_{\rm out}^\dag(\omega)+{\cal R}_{d}(\omega)\hat{d}_{\rm out}^\dag(\omega)+\sum_i{\cal S}_{d,i}(\omega)\hat{A}_{{\rm out},i}^\dag(\omega) ,\nonumber\\
\label{outdamp2}
\eqa 
where $|{\cal R}_{s,d}(\omega)|^2+|{\cal T}_{s,d}(\omega)|^2+\sum_i|{\cal S}_{s,d, i}(\omega)|^2=1$. 

\renewcommand{\theequation}{F-\arabic{equation}}
\setcounter{equation}{0}  
\section*{APPENDIX F}  
Here we provide the analytical forms for the ${\cal T}(\omega)$'s for $n=1,~2,~3,~5$ and 7 nanoparticles respectively with damping, where we have set $g_{\rm out}=g_{\rm in}$,
\bqa
{\cal T}_1(\omega)&=&\frac{2g_{\rm in}}{2g_{\rm in}+\Gamma-2i (\omega-\omega_0)}, \nonumber \\
{\cal T}_2(\omega)&=&\frac{-4ig_{\rm np} g_{\rm in}}{4g_{\rm np}^2+(g_{\rm in}+\Gamma-2i (\omega-\omega_0))^2}, \nonumber \\
{\cal T}_3(\omega)&=&-8g_{\rm np}^2 g_{\rm in}[(g_{\rm in}+\Gamma-2i (\omega-\omega_0))\times  \nonumber \\
&& \hskip-0.2cm (8g_{\rm np}^2+(\Gamma-2i(\omega-\omega_0))(g_{\rm in}+\Gamma-2i(\omega-\omega_0)))]^{-1}, \nonumber \\
{\cal T}_5(\omega)&=&32g_{\rm np}^4 g_{\rm in}[(4g_{\rm np}^2+(\Gamma+2i(\omega-\omega_0))   \times        \nonumber \\
&&\qquad \qquad(g_{\rm in}+\Gamma-2i(\omega-\omega_0)))\times   \nonumber \\
&& \qquad \qquad (4g_{\rm np}^2(2 g_{\rm in}+3(\Gamma-2i(\omega-\omega_0)))   \nonumber \\
&& ~~+ (\Gamma-2i(\omega-\omega_0))^2(g_{\rm in}+\Gamma-2i(\omega-\omega_0)))]^{-1}, \nonumber
\eqa 
and
\bqa
{\cal T}_7(\omega)&=&-128g_{\rm np}^6 g_{\rm in}[((4g_{\rm np}^2(g_{\rm in}+2(\Gamma -2i(\omega-\omega_0)))+ \nonumber \\
&&\qquad(\Gamma-2i(\omega-\omega_0))^2(g_{\rm in}+\Gamma-2i(\omega-\omega_0))) \times \nonumber \\
&&\qquad~(32g_{\rm np}^4+4g_{\rm np}^2(3 g_{\rm in}+4(\Gamma-2i(\omega-\omega_0))) \times \nonumber \\
&&\qquad ~~~~ (\Gamma - 2i(\omega-\omega_0))+(\Gamma-2i(\omega-\omega_0))^3 \times \nonumber \\
&&\qquad \qquad \qquad \qquad(g_{\rm in}+\Gamma-2 i (\omega-\omega_0))))]^{-1}. \nonumber 
\eqa

\renewcommand{\theequation}{G-\arabic{equation}}
\setcounter{equation}{0}  
\section*{APPENDIX G}  
In our discussion of two-plasmon interference we needed to evaluate the probabilities for finding plasmons in the output state given in Eq.~(\ref{outputstate}). For an arbitrary state $\ket{\psi}$, the probability of detecting one plasmon in a given mode at any frequency $\omega$ is given by $P(1)=\int_{-\infty}^{\infty}{\rm d}\omega |\braket{1_\omega}{\psi}|^2$, where $\ket{n_\omega}$ is the continuous mode number state, as previously defined. The probability of detecting two plasmons in a given mode, one at any frequency $\omega$ and the other at any frequency $\omega'$ is then $P(2)=\int_{-\infty}^{\infty}{\rm d}\omega \int_{-\infty}^{\infty}{\rm d}\omega' |\braket{2_{(\omega,\omega')}}{\psi}|^2$, where $\ket{2_{(\omega,\omega')}}=\frac{1}{\sqrt{2}} \hat{s}^\dag_{\rm out}(\omega) \hat{s}^\dag_{\rm out}(\omega')\ket{0}$ is the continuous mode pair-state~\cite{Loudon}, which allows for each plasmon to have a different frequency profile. Thus we have the following probabilities
\bqa
P(2_{s}, 0_{d}) &=& \frac{1}{2} \int_{-\infty}^{\infty} {\rm d}\omega \int_{-\infty}^{\infty} {\rm d}\omega' \times \nonumber \\
&&\qquad \qquad \qquad |~ _{s,d,A}\bra{0} \hat{s}_{\rm out}(\omega') \hat{s}_{\rm out}(\omega) \ket{\psi}_{\rm out}|^{2},\nonumber\\
P(0_{s}, 2_{d}) &=& \frac{1}{2} \int_{-\infty}^{\infty} {\rm d}\omega \int_{-\infty}^{\infty} {\rm d}\omega' \times \nonumber\\
&&\qquad \qquad \qquad |~ _{s,d,A} \bra{0} \hat{d}_{\rm out}(\omega') \hat{d}_{\rm out}(\omega)\ket{\psi}_{\rm out}|^{2} \nonumber \\
P(1_{s}, 1_{d}) &=& \int_{-\infty}^{\infty} {\rm d}\omega \int_{-\infty}^{\infty} {\rm d}\omega'  \times \nonumber\\
&&\qquad \qquad \qquad |~ _{s,d,A} \bra{0} \hat{s}_{\rm out}(\omega) \hat{d}_{\rm out}(\omega') \ket{\psi}_{\rm out}|^{2}, \nonumber \\
P(1_{s}, 0_{d}) &=& \int_{-\infty}^{\infty} {\rm d}\omega |~ _{s,d,A} \bra{0} \hat{s}_{\rm out}(\omega) \ket{\psi}_{\rm out}|^{2},\nonumber\\
P(0_{s}, 1_{d}) &=& \int_{-\infty}^{\infty} {\rm d}\omega |~ _{s,d,A} \bra{0} \hat{d}_{\rm out}(\omega) \ket{\psi}_{\rm out}|^{2},\nonumber\\
P(0_{s}, 0_{d}) &=& |~ _{s,d,A} \left< 0 | \psi \right>_{\rm out}|^{2} .\nonumber
\eqa
These may be evaluated by using the relationship between the input and output field operators given in Eq.~(\ref{outdamp2}) and the following commutation relations
\bqa
\left[\hat{s}_{\rm out}(\omega),\hat{s}_{\rm out}^{\dagger}(\omega) \right] &=&\delta(\omega-\omega') =\left[\hat{d}_{\rm out}(\omega),\hat{d}_{\rm out}^{\dagger}(\omega) \right],\nonumber\\
\left[\hat{s}_{\rm out}(\omega),\hat{d}_{\rm out}^{\dagger}(\omega) \right] &=& \left[\hat{d}_{\rm out}(\omega),\hat{s}_{\rm out}^{\dagger}(\omega) \right] =0, \nonumber\\
\left[\hat{s}_{\rm out}(\omega),\hat{F}_{s}^{\dagger}(\omega) \right] &=& \left[\hat{s}_{\rm out}(\omega),\hat{F}_{d}^{\dagger}(\omega) \right]  \nonumber\\ 
&=& \left[\hat{s}_{\rm out}(\omega),\hat{F}_{s}(\omega) \right]  \nonumber\\ 
&=& \left[\hat{s}_{\rm out}(\omega),\hat{F}_{d}(\omega) \right] =0,  \nonumber
\eqa
and
\bqa
\left[ \hat{F}_{s}(\omega),\hat{F}_{s}^{\dagger}(\omega') \right] &=&\delta(\omega-\omega') (1-|{\cal R}_{s}(\omega)|^{2}-|{\cal T}_{s}(\omega)|^{2}),\nonumber\\
\left[\hat{F}_{d}(\omega),\hat{F}_{d}^{\dagger}(\omega') \right] &=&\delta(\omega-\omega') (1-|{\cal R}_{d}(\omega)|^{2}-|{\cal T}_{d}(\omega)|^{2}),\nonumber\\
\left[\hat{F}_{s}(\omega),\hat{F}_{d}^{\dagger}(\omega') \right] &=& -\delta(\omega-\omega')({\cal R}_{s}^{*}(\omega){\cal T}_{d}(\omega)+{\cal T}_{s}^{*}(\omega){\cal R}_{d}(\omega)), \nonumber\\
\left[\hat{F}_{d}(\omega),\hat{F}_{s}^{\dagger}(\omega') \right] &=& -\delta(\omega-\omega')({\cal T}_{d}^{*}(\omega){\cal R}_{s}(\omega)+{\cal R}_{d}^{*}(\omega){\cal T}_{s}(\omega)). \nonumber
\eqa

By straightforward substitution, one finds
\begin{widetext}
\bqa
P(2_{s}, 0_{d}) &=& \frac{1}{2} \int_{-\infty}^{\infty}{\rm d}\omega \int_{-\infty}^{\infty}{\rm d} \omega' | \psi(\omega,\omega') {\cal R}_{s}(\omega) {\cal T}_{d}(\omega')+\psi(\omega',\omega) {\cal R}_{s}(\omega') {\cal T}_{d}(\omega) |^{2}, \nonumber\\
P(2_{s}, 0_{d}) &=& \frac{1}{2} \int_{-\infty}^{\infty}{\rm d}\omega \int_{-\infty}^{\infty}{\rm d} \omega' | \psi(\omega,\omega') {\cal T}_{s}(\omega) {\cal R}_{d}(\omega')+\psi(\omega',\omega) {\cal T}_{s}(\omega') {\cal R}_{d}(\omega) |^{2},  \nonumber\\
P(1_{s}, 1_{d}) &=&  \int_{-\infty}^{\infty}{\rm d}\omega \int_{-\infty}^{\infty}{\rm d}\omega' | \psi(\omega,\omega') {\cal R}_{s}(\omega) {\cal R}_{d}(\omega')+\psi(\omega',\omega) {\cal T}_{s}(\omega') {\cal T}_{d}(\omega)|^{2}, \nonumber\\
P(1_{s}, 0_{d}) &=& \int_{-\infty}^{\infty} {\rm d}\omega \int_{-\infty}^{\infty} {\rm d}\omega' \bigg\{ |\psi(\omega,\omega')|^{2} |{\cal R}_{s}(\omega)|^{2}(1-|{\cal R}_{d}(\omega')|^{2}-|{\cal T}_{d}(\omega')|^{2}) \nonumber\\
&&~~~~~~~~~~~~~~~~~~~~~~~~+ |\psi(\omega',\omega)|^{2} |{\cal T}_{d}(\omega)|^{2}(1-|{\cal R}_{s}(\omega')|^{2}-|{\cal T}_{s}(\omega')|^{2}) \nonumber\\
&&~~~~~~~~~~~~~~~~~~~~~~~~~~~~~~~~ -\psi^{*}(\omega,\omega') \psi(\omega',\omega) {\cal R}_{s}^{*}(\omega){\cal T}_{d}(\omega) ({\cal T}_{d}^{*}(\omega'){\cal R}_{s}(\omega')+{\cal R}_{d}^{*}(\omega'){\cal T}_{s}(\omega'))  \nonumber\\ 
&&~~~~~~~~~~~~~~~~~~~~~~~~~~~~~~~~~~~~~~~~-\psi^{*}(\omega',\omega) \psi(\omega,\omega') {\cal T}_{d}^{*}(\omega){\cal R}_{s}(\omega) ({\cal R}_{s}^{*}(\omega'){\cal T}_{d}(\omega')+{\cal T}_{s}^{*}(\omega'){\cal R}_{d}(\omega')) \bigg\}, \nonumber\\ 
P(0_{s}, 1_{d}) &=& \int_{-\infty}^{\infty} {\rm d}\omega \int_{-\infty}^{\infty} {\rm d}\omega' \bigg\{ |\psi(\omega,\omega')|^{2} |{\cal T}_{s}(\omega)|^{2}(1-|{\cal R}_{d}(\omega')|^{2}-|{\cal T}_{d}(\omega')|^{2})\nonumber\\
&&~~~~~~~~~~~~~~~~~~~~~~~~+ |\psi(\omega',\omega)|^{2} |{\cal R}_{d}(\omega)|^{2}(1-|{\cal R}_{s}(\omega')|^{2}-|{\cal T}_{s}(\omega')|^{2}) \nonumber\\
&&~~~~~~~~~~~~~~~~~~~~~~~~~~~~~~~~ -\psi^{*}(\omega,\omega') \psi(\omega',\omega) {\cal T}_{s}^{*}(\omega){\cal R}_{d}(\omega) ({\cal T}_{d}^{*}(\omega'){\cal R}_{s}(\omega')+{\cal R}_{d}^{*}(\omega'){\cal T}_{s}(\omega'))  \nonumber\\ 
&&~~~~~~~~~~~~~~~~~~~~~~~~~~~~~~~~~~~~~~~~-\psi^{*}(\omega',\omega) \psi(\omega,\omega') {\cal R}_{d}^{*}(\omega){\cal T}_{s}(\omega) ({\cal R}_{s}^{*}(\omega'){\cal T}_{d}(\omega')+{\cal T}_{s}^{*}(\omega'){\cal R}_{d}(\omega')) \bigg\}, \nonumber\\ 
P(0_{s}, 0_{d}) &=& \int_{-\infty}^{\infty} {\rm d}\omega \int_{-\infty}^{\infty} {\rm d}\omega' \bigg\{ |\psi(\omega,\omega')|^{2} (1-|{\cal R}_{s}(\omega)|^{2}-|{\cal T}_{s}(\omega)|^{2})(1-|{\cal R}_{d}(\omega')|^{2}-|{\cal T}_{d}(\omega')|^{2})\nonumber\\
&&~~~~~~~~~~~~~~~~~~~~~~~~+\psi^{*}(\omega',\omega) \psi(\omega,\omega') ({\cal T}_{d}^{*}(\omega'){\cal R}_{s}(\omega')+{\cal R}_{d}^{*}(\omega'){\cal T}_{s}(\omega'))({\cal R}_{s}^{*}(\omega'){\cal T}_{d}(\omega')+{\cal T}_{s}^{*}(\omega'){\cal R}_{d}(\omega')) \bigg\} ,\nonumber
\eqa
\end{widetext}
where only terms making a nonzero contribution have been retained. Considering the case where the transmission coefficients are approximately constant over the range of frequencies for which $|\psi(\omega, \omega')|$ is significant one finds
\bqa
P(2_{s}, 0_{d}) &\approx& |{\cal R}_{s} |^{2} |{\cal T}_{d}|^{2} (1+ {\cal I}), \nonumber\\
P(0_{s}, 2_{d}) &\approx& |{\cal T}_{s} |^{2} |{\cal R}_{d}|^{2} (1+ {\cal I}), \nonumber\\
P(1_{s}, 1_{d}) &\approx& |{\cal R}_{s}|^{2} |{\cal R}_{d}|^{2}+ |{\cal T}_{s}|^{2} |{\cal T}_{d}|^{2} \nonumber\\
&&+({\cal R}_{s}{\cal R}_{d}{\cal T}_{s}^{*}{\cal T}_{d}^{*}+{\cal R}_{s}^{*}{\cal R}_{d}^{*}{\cal T}_{s}{\cal T}_{d}){\cal I}, \nonumber\\
P(1_{s}, 0_{d}) &\approx& |{\cal R}_{s}|^{2}(1-|{\cal R}_{d}|^{2}-|{\cal T}_{d}|^{2}) \nonumber\\
&& +|{\cal T}_{d}|^{2}(1-|{\cal R}_{s}|^{2}-|{\cal T}_{s}|^{2}) \nonumber \\
&&-(2|{\cal R}_{s}|^{2}|{\cal T}_{d}|^{2}+{\cal R}_{s}^{*}{\cal R}_{d}^{*}{\cal T}_{s}{\cal T}_{d}+{\cal R}_{s}{\cal R}_{d}{\cal T}_{s}^{*}{\cal T}_{d}^{*}){\cal I}, \nonumber\\
P(0_{s}, 1_{d}) &\approx& |{\cal T}_{s}|^{2}(1-|{\cal R}_{d}|^{2}-|{\cal T}_{d}|^{2}) \nonumber\\
&& +|{\cal R}_{d}|^{2}(1-|{\cal R}_{s}|^{2}-|{\cal T}_{s}|^{2}) \nonumber \\
&&-(2|{\cal R}_{d}|^{2}|{\cal T}_{s}|^{2}+{\cal T}_{s}^{*}{\cal T}_{d}^{*}{\cal R}_{s}{\cal R}_{d}+{\cal T}_{s}{\cal T}_{d}{\cal R}_{s}^{*}{\cal R}_{d}^{*}){\cal I}, \nonumber\\
P(0_{s}, 0_{d}) &\approx& (1-|{\cal R}_{s}|^{2}-|{\cal T}_{s}|^{2})(1-|{\cal R}_{d}|^{2}-|{\cal T}_{d}|^{2})  \nonumber\\
&&+ ({\cal T}_{d}^{*}{\cal R}_{s}+{\cal R}_{d}^{*}{\cal T}_{s})({\cal R}_{s}^{*}{\cal T}_{d}+{\cal T}_{s}^{*}{\cal T}_{d}){\cal I}, \nonumber
\eqa
where we have introduced the (real) overlap integral
\be
{\cal I}=\int_{-\infty}^{\infty} {\rm d}\omega \int_{-\infty}^{\infty} {\rm d}\omega' \psi(\omega,\omega') \psi^{*}(\omega',\omega).\nonumber
\ee


\end{document}